\documentclass[%
 reprint,
superscriptaddress,
 amsmath,amssymb,
aip, jcp,
citeautoscript]{revtex4-2}

\usepackage{graphicx}
\usepackage{dcolumn}
\usepackage{bm}
\usepackage{braket}
\usepackage{hyperref}

\usepackage{lipsum}
\usepackage{xcolor}

\everymath{\medmuskip=1.5mu minus 1.5mu\thickmuskip=2mu minus 2mu}

\newcommand{\DIPC}[0]{
Donostia International Physics Center (DIPC),
Paseo Manuel de Lardizabal 4, 20018 Donostia-San Sebasti\'an, Spain}
\newcommand{\CFM}[0]{
Centro de F\'{\i}sica de Materiales CFM/MPC (CSIC-UPV/EHU), Paseo Manuel de Lardizabal 5, 20018 Donostia-San Sebasti\'an, Spain}
\newcommand{\PolymerEHU}[0]{Departamento de Pol\'{i}meros y Materiales Avanzados: F\'{i}sica, Qu\'{i}mica y Tecnolog\'{i}a, Facultad de Qu\'{i}micas (UPV/EHU), Apartado 1072, 20080 Donostia-San Sebasti\'{a}n, Spain}

\newcommand{\burdeos}[0]{Institut des Sciences Moléculaires (ISM), Université de Bordeaux, 351 Cours de la Libération, 33405 Talence, France}

\newcommand{\UAM}[0]{Departamento de Química Física Aplicada, Universidad Autónoma de Madrid, 28049, Madrid, Spain}

\begin{document}

\preprint{APS/123-QED}

\title{Asymmetry and coverage dependence in two-pulse correlation measurements of CO photodesorption from Pd(111): Insights from theory}

\author{Raúl Bombín}
\email{raul.bombin@u-bordeaux.fr}
\affiliation{\burdeos}
\affiliation{\DIPC}

\author{Alberto S. Muzas}
\email{alberto.muzas@uam.es}
\affiliation{\UAM}
\email{alberto.muzas@uam.es}

\author{Alfredo Serrano Jiménez}
\affiliation{\PolymerEHU}
\affiliation{\DIPC}
\affiliation{\CFM}

\author{J. I\~{n}aki Juaristi}
\email{josebainaki.juaristi@ehu.eus}
\affiliation{\PolymerEHU}
\affiliation{\CFM}
\affiliation{\DIPC}
\author{Maite Alducin}
\email{maite.alducin@ehu.eus}
\affiliation{\CFM}
\affiliation{\DIPC}

\date{\today}

\begin{abstract} 
Two-pulse correlation experiments  performed using pulses of different intensities on Pd(111) with different CO coverages showed that the CO photodesorption probability depends on whether the strong or the weak pulse arrives first to the surface, being this difference particularly large for the low-covered surface. Motivated by these experiments, we perform molecular dynamics simulations using a multicoverage potential energy surface that was previously constructed with the embedded atom neural network method. The process is modeled by combining the two-temperature model (2TM)--to describe the laser-excited electrons and phonons--and Langevin dynamics with electronic time-dependent temperature $T_\textrm{e}$--to model the coupling of the nuclei degrees of freedom with the laser-excited electrons. 
We show that improving the energy balance description in 2TM--by including \textit{ab-initio} $T_\textrm{e}$-dependent electronic heat capacity and electron-phonon coupling constant--is key to reproduce the asymmetry of the photodesorption probability $P_{\rm des}$ between positive and negative time delays. Furthermore, we also explore possible reasons for the usual underestimation of $P_{\rm des}$ at zero delay given by state-of-the-art calculations. In particular, we improve the description of the energy exchange between CO and the metal surface at high $T_\textrm{e}$ by 
including in the simulations a $T_\textrm{e}$-dependent friction coefficient. 
The prediction for $P_{\rm des}$ at zero delay in this case, increases by an order of magnitude, reducing its discrepancy with the experimental value. Altogether, our results hint to the importance of accounting for the temperature dependence of the electronic structure and properties in describing the extreme conditions generated in 2PC experiments.
\end{abstract}

\maketitle

\section{\label{sec:intro} Introduction}

The use of intense femtosecond laser pulses in the visible and near-infrared regime constitutes and efficient way to induce surface reactions. In this wavelength range, initially the laser pulse excites the electronic system. Subsequently, the excited electrons can transfer energy to the adsorbates and the surface phonons. As a result, new reaction pathways different from those found in thermal reactions are activated, leading to larger reaction cross sections and even new reactions~\cite{frischkorncr06,saalfrank2006,Guo99}. 

In the process of understanding these experiments, a natural and important question arises. Since the reacting adsorbates evolve in a highly excited system, consisting of electrons excited directly by the laser pulse and phonons excited by the decay of such excited electrons, it is important to disentangle whether the observed reactions are governed by the excited electrons or the excited phonons. Experimentally, information about this matter can be obtained by performing 
two-pulse correlation (2PC) experiments. In these experiments, the laser induced adsorbate desorption yield is measured as a function of the time delay between two subsequent laser pulses. More precisely, a correlation time is defined as the full-width-at-half-maximum (FWHW) of the desorption probability curve as a function of time delay. The information that can be obtained from these experiments relays in the different time scales of the different excitations. Specifically, the electrons excited by the laser pulse are driven out of equilibrium with the surface lattice for times of around 1-2 ps.  For this reason, when the measured correlation time is of the order of one ps, the process is typically deemed to be electron mediated. Likewise, correlation times beyond 20 ps are considered to be a clear signature of a phonon-mediated process. Still, intermediate correlation times between these values must be considered ambiguous because both electronic and phononic systems can be excited and their relative importance for the particular reaction will depend on the strength of the coupling of the adsorbates to each of the two subsystems. 

Several adsorbate/substrate systems have been investigated using the 2PC technique. In most of the studied systems that include, photodesorption in NO/Pd(111)~\cite{Budde91}, O$_2$/Pt(111)~\cite{Kao93,Deliwala95}, O$_2$/Pd(111)~\cite{Misewich94}, CO/Cu(100)~\cite{Struck96}, NO/Pt(111)~\cite{yamanaka2002}, and recombinative photodesorption of H$_2$ from H+H/Ru(0001)~\cite{Denzler2003}, correlation times in the range 1-3 ps were reported and ascribed to an electronically governed process. Only the CO desorption from Ru(0001), both when adsorbed molecularly~\cite{Funk2000,Gladh2013} and when it requires the recombination of C and O adsorbates~\cite{Wagner2008}, has been considered to be unquestionably governed by phonons, with correlation times of around 20 ps. Interestingly, in the O+CO/Ru(0001) system, the correlation times for CO oxidation and CO desorption are 3 ps and 20 ps, respectively~\cite{Bonn99,Oberg2015}. Therefore, CO oxidation was interpreted as an electronically driven mechanism and CO desorption as a phonon driven mechanism.  

Another system that has been the subject of extensive experimental analysis is the photodesorption of CO from Pd(111)~\cite{Szymanski2007,Hong2016}. In this case, special attention was paid to the dependence of the measured quantities on adsorbate coverage. More precisely, aimed to determine whether electrons or phonons dominate the photoinduced desorption of CO from Pd(111) and its dependence on coverage, Hong \textit{et al.}~\cite{Hong2016} performed 2PC measurements using two laser pulses of 780~nm, FWHM of 130~fs and $\mathrm{sech^{2}}$ profile, but different fluence, namely, a strong p-polarized pulse with fluence 2.4 times higher than the weaker s-polarized pulse. Positive (negative) time delays correspond to the weak pulse arriving before (after) the strong pulse. Considering that the desorption probability increases in about two orders of magnitude when varying the coverage from 0.33~ML to 0.75~ML, also the laser fluences were chosen differently for each coverage in order to obtain similar desorption probabilities at zero delay. The measured 2PC desorption probabilities as a function of time delay showed different half-width-at-half-maximum (HWHM) between positive (HWHM$^+$) and negative (HWHM$^-$) delays and among the investigated coverages. Regarding the numerical values, only the HWHM$^-$ of 1.1~ps that was obtained at negative delays for 0.24~ML could be consistent with electron-dominated desorption. In the rest of cases, the HWHM values varied between 6.7 and 25.5~ps. Altogether, the 2PC measurements were unable to clarify whether electrons or phonons dominate the CO photodesorption from Pd(111) and, particularly, the reasons for the asymmetric HWHM between positive and negative delays and its dependence on coverage. 

Motivated by these results and in order to disentangle the mechanisms governing the observed behavior, we present a study based on molecular dynamics simulations aimed to reproduce and understand the 2PC experiments in this system. The 2PC simulations are performed for two different coverages, namely, 0.33~ML and 0.75~ML, using our more recent implementation to treat the femtosecond laser pulse induced dynamics at metal surfaces~\cite{Muzas2024}. In the simulations, which incorporate the coupling to the laser excited electrons of both the adsorbate and surface atoms, the adiabatic forces are calculated with a multidimensional neural network (NN) potential energy surface (PES) trained with \textit{ab-initio} data, which is valid and accurate for multiple CO coverages at the Pd(111) surface~\cite{Muzas2024}. 
As a methodological novelty, temperature-dependent electronic heat capacity and electron-phonon coupling constant based on \textit{ab-initio} calculations~\cite{Li2022} are included in the energy balance equations between the electron and phonon baths. The results of our simulations based on the improved electronic temperature $T_\textrm{e}$ provide a better description of the measured asymmetry between positive and negative time delays and its dependence on adsorbate coverage. As an additional methodological novelty, we calculate the temperature dependence of the adsorbate-electron coupling in the CO/Pd(111) system and find that its effect in the CO desorption probability becomes increasingly important at experimental conditions for which $T_\textrm{e}\geq 5000$~K. 

The paper is organized as follows. Section~\ref{sec:theory} describes the theoretical model that is used to simulate the femtosecond laser induced dynamics, including the two different parameterizations of the two-temperature model (2TM) that we employ to describe the energy balance between electrons and phonons in Pd(111) and also the calculation of the temperature-dependent friction coefficient for CO. In Sec.~\ref{sec:results}, we present and compare the results of our different simulations to the 2PC measurements, discussing the improvements of our implementation compared to the current state-of-the-art. Finally, Sec.~\ref{sec:summary} summarizes the main conclusions of our work.

\section{\label{sec:theory} Theoretical model}
\subsection{\label{sec:md} Machine Learning Molecular Dynamics}
The dynamics of adsorbates that are induced by intense femtosecond laser pulses can be described as a two-step process in which the laser first excites the metal electrons and, next, those excited electrons couple to the surface (phonons) and adsorbate degrees of freedom. The rapid decay of the initial excited electrons into a thermal Fermi-Dirac distribution in metals justifies the use of the two-temperature model~\cite{Anisimov1974} to describe the perturbation created by the laser pulse in terms of coupled electronic and phononic thermal baths~\cite{Budde91,Kao93,Struck96,frischkorncr06,saalfrank2006,caruso22}. Next, the coupling of the excited electronic bath with the adsorbate dynamics is commonly modeled through Langevin equations of motion~\cite{Springer1994, Springer1996, Luntz2006, Vazhappilly2009, Fuchsel2011,juaristiprb17, Lindner2023}. Also the effect of the excited surface phonons has been progressively incorporated in most subsequent studies using different kinds of thermostats to describe the surface phononic bath~\cite{loncaricprb16, loncaricnimb16, Scholz2016, alducin2019, Scholz2019, tetenoire2022, tetenoire2023, Zugec2024, Gonzalez2025, Mladineo2025, Wang2025}.

All the MD simulations presented in this work are performed following our most recent and consistent implementation [denoted ($T_\textrm{e}$,$T_\textrm{l}$)-MDEF in the following], in which the effect of the excited electrons is directly incorporated not only in the adsorbate dynamics but also in the lattice dynamics~\cite{Muzas2024}. In particular, the motion of each atom conforming the adlayer and mobile atoms of the metal surface is described by the following Langevin equation:
\begin{eqnarray}\label{eq:langevin}
m_i\frac{\text{d}^2\mathbf{r}_i}{\text{d}t^2}&=&-\nabla_{\mathbf{r}_i} 
V(\mathbf{r}_1,...,\mathbf{r}_N)-\eta_{e,i}(\mathbf{r}_i)\frac{\text{d}\mathbf{r}_i}{\text{d}t} \\ \nonumber
& & + \mathbf{R}_{e,i}[T_{e}(t),\eta_{e,i}(\mathbf{r}_i)] \, , 
\end{eqnarray}
where $m_i$, $\textbf{r}_i$, and $\eta_{e,i}(\mathbf{r}_i)$ are the mass, position vector, and electronic friction coefficient of the $i^{th}$ atom in the system. The adiabatic force [first term in the right hand side of Eq.~\eqref{eq:langevin}] is here calculated with the multicoverage NN PES developed in Ref.~\onlinecite{Muzas2024}, that was trained with data sets of energies and forces that were extracted from ($T_\textrm{e}$,$T_\textrm{l}$)-AIMDEF simulations~\cite{alducin2019} based on density functional theory and the van der Waals exchange-correlation functional developed by Dion \textit{et al}~\cite{Dion2004}. As shown there, the constructed adiabatic multidimensional and multicoverage NN-PES reproduces well for arbitrary atomic positions $(\mathbf{r}_1,...,\mathbf{r}_N)$,  the ground state energy of the Pd(111) surface with a CO overlayer whose coverage can vary from 0.33~ML to 0.75~ML. The effect of the electronic excitations and deexcitations on the adsorbate and lattice dynamics is described by the electronic friction and electronic stochastic forces [second and third terms in Eq.~\eqref{eq:langevin}, respectively], which are related  through the fluctuation-dissipation theorem. In particular, $\mathbf{R}_{e,i}$ is modeled by a Gaussian white noise with variance $ 
\mathrm{Var}[\mathbf{R}_{\text{e},i}(T_\text{e},\eta_{\text{e},i})]=[2 k_\text{B} T_\text{e}(t) \eta_{\text{e},i}(\mathbf{r}_i)]/\Delta t$,  where $k_\text{B}$, $\Delta t$, and $T_\text{e}(t)$ are the Boltzmann constant, time-integration step, and time-dependent temperature of the laser-excited electronic bath, respectively. This temperature is obtained from an independent calculation using a two-temperature model as described in Sec.~\ref{2TM}.

The electronic friction coefficient for each atom in the adlayer (C and O) is calculated within the local density friction approximation (LDFA)~\cite{juaristi08,alducinpss17} as the friction coefficient that the atom under consideration would experience when embedded in an homogeneous free electron gas with density equal to the value of the bare metal surface electronic density at the position $\textbf{r}_i$ where the adlayer atom is located at each instant. In particular, the electronic density is evaluated at each integration step using the surface-density-generator function proposed in Ref.~\onlinecite{Serrano2021}. Let us remark at this point that the physically motivated density that, in principle, has to be included is the bare metal electronic density because the LDFA friction coefficient already encloses the information of the perturbation created by the embedded atom in the metal electron density. Note also that this perturbation also includes information of the embedded atom orbitals, as evidenced by the ability to reproduce the $Z-$oscillations of the stopping-power measured in ion-solid collision~\cite{winter2003}.

In the case of Pd atoms, the friction coefficient is calculated in terms of its embedding electronic density $\rho(\mathbf{r}_i)$, i.e., the density created by the rest of Pd atoms at its position as:
\begin{equation}
\label{eq:friccoef-den1}
    \eta_i^\mathrm{Pd}(\mathbf{r}_i)=\left(\frac{r_\mathrm{s}^0}{r_\mathrm{s}(\mathbf{r}_i)}\right)^3\eta^\mathrm{Pd}(r_\mathrm{s}^0) 
\end{equation}
with
\begin{equation}
\label{eq:friccoef-den2}
  \eta^\mathrm{Pd}(r_\mathrm{s}^0)=\frac{m_\mathrm{Pd}}{3k_\mathrm{B}d_{\mathrm{Pd}}}G \, .   
\end{equation}
In these equations, $r_s(\mathbf{r}_i)$ is the Wigner-Seitz radius associated to the embedding electronic density $\rho(\mathbf{r}_i)$, $r_s^0=4.38$~a.u. (atomic units) is the Weigner-Seitz radius associated to the calculated embedding density of one Pd atom at its equilibrium position in the bulk, $\eta^\mathrm{Pd}(r_\mathrm{s}^0)$ is the Pd friction coefficient at $r_\mathrm{s}^0$, $m_\mathrm{Pd}$ is the mass of one Pd atom, $d_\mathrm{Pd}$ is the atomic density of Pd (4 atoms per 64 \AA$^3$ according to our bulk model), and $G$ is the Pd electron-phonon energy exchange coupling factor. For those calculations in which $G$ is kept constant (2TM-1, see section \ref{2TM}), $\eta^\mathrm{Pd}(r_\mathrm{s}^0)$ is constant and equal to 1.6223 a.u. as in Ref.~\onlinecite{Muzas2024}. When $G$ is allowed to depend on $T_\mathrm{e}$ (2TM-2, see section \ref{2TM}), then the following polynomial fit
\begin{equation}
    \label{eq:G_Te}
G(T_\mathrm{e})=G(0)\left(1+\sum_{k=1}^{4}b_k T_\mathrm{e}^k\right) \, ,
\end{equation}
with $G(0)=7.87\times$10$^{17}$ WK$^{-1}$m$^{-3}$, is used to evaluate $\eta^\mathrm{Pd}(r_\mathrm{s}^0,T_\mathrm{e})$ during dynamics. The coefficients of this fit are shown in Table~\ref{tab:fiteta}.

\subsection{\label{2TM}Two-Temperature Model Applied to Palladium}

Following pioneering works~\cite{Corkum1988, Prybyla1992}, the standard 2TM equations describing the coupling of the electronic and phononic thermal baths created by intense femtosecond laser pulses read,
\begin{equation}
\begin{split}\label{eq:2TM}
    C_\mathrm{e}\frac{\partial T_\mathrm{e}}{\partial t} &=
        \frac{\partial}{\partial z}\left(\kappa_\mathrm{e}\frac{\partial T_\mathrm{e}}{\partial z}\right)-G(T_\mathrm{e}-T_\mathrm{l})+S(z,t)\\
    C_\mathrm{l}\frac{\partial T_\mathrm{l}}{\partial t} &=
        G(T_\mathrm{e}-T_\mathrm{l}),
\end{split}
\end{equation}
where $C_\mathrm{e}$ and $C_\mathrm{l}$ are the electron and lattice heat capacities, respectively; $\kappa_\mathrm{e}$ is the electron thermal conductivity, whose dependence on the electronic and phononic temperatures ($T_\textrm{e}$ and  $T_\textrm{l}$) can be reasonably approximated by~\cite{Ashcroft1988}, $\kappa_\mathrm{e}(T_\mathrm{e},T_\mathrm{l})=\kappa_0 T_\mathrm{e}/T_\mathrm{l}$, where $\kappa_0=72$~WK$^{-1}$m$^{-1}$ is the Pd thermal conductivity at 300~K~\cite{Kittel1986}; 
$G$ is the electron-phonon energy exchange coupling constant; and $S(z,t)$ is the laser heating source. In using Eq.~\eqref{eq:2TM}, we are assuming that the energy of the laser pulse is absorbed uniquely by the electron bath, with the absorbed fluence being constant in the plane parallel to the metal surface, and that lattice thermal diffusion into the bulk is negligible in comparison to the electronic one (first right hand side term in the first equation) for the time scales of interest~\cite{frischkorncr06}. 

The 2TM equations are solved for the specificities of the system and experiments under consideration. Thus, following the experimental work by Szymanski \textit{et al.}~\cite{Szymanski2007} , the laser source term describing two delayed pulses with absorbed fluences $F_a$ and $F_b$ that arrive at times $t_a$ and $t_b$, respectively, is modeled as,
\begin{align}
    S(z,t)= & \frac{\alpha_\lambda}{2\sigma} \mathrm{e}^{-\alpha_\lambda z} \nonumber\\
     & \times\left[ F_a \, 
    \mathrm{sech}^2\left(\frac{t-t_a}{\sigma}\right) +  F_b \, 
    \mathrm{sech}^2\left(\frac{t-t_b}{\sigma}\right)\right ].
\end{align}
In this equation, $\sigma=0.5\,\mathrm{FWHM}/\mathrm{arccosh}{\sqrt{2}}$ and $\alpha_\lambda$ is the substrate absorption coefficient for light wavelength $\lambda$, which is obtained from the extinction coefficient $k_{\lambda}$ as $\alpha_\lambda=(4\pi k_\lambda)/\lambda$. In the case of Pd, $k_{\lambda}=4.986$ at 780~nm~\cite{Szymanski2007}.

The temperature dependence of $C_\mathrm{l}$ for Pd(111) is described using the fitting expression developed by Szymanski~\textit{et al.}~\cite{Szymanski2007} that reads,
\begin{equation}
    C_\mathrm{l}(T_\mathrm{l}) = 
    C_{2}+\frac{C_{1}-C_{2}}{1+(T_\mathrm{l}/T_{0})^p}+mT_\mathrm{l} \, ,
\end{equation}
where $C_{1}=-3.577785\times10^{-5}$~JK$^{-1}$m$^{-3}$, $C_{2}=2.801128\times10^{6}$~JK$^{-1}$m$^{-3}$, $T_{0}=62.98191$~K, $p=2.06271$, and $m=278.44328$~JK$^{-2}$m$^{-3}$ are the optimized fitting coefficients that reproduce the experimental $C_\mathrm{l}$ values compiled in Ref.~\onlinecite{Miiller1971} for this surface at temperatures below 1000~K. The fitted $C_\mathrm{l}(T_\mathrm{l})$ expression is expected to work well in our case because, as shown below, the highest values of the phononic temperatures calculated with 2TM for the 2PC experimental conditions remain close to this upper limit of 1000~K. 

Regarding the dependence of $C_\mathrm{e}$ and $G$ on $T_\textrm{e}$, the electronic temperature can reach values well above the melting point of the metal substrate, making it difficult to measure these quantities within the temperature range of interest (up to 8000~K). In fact, to the authors knowledge, there are no experimental measurements of the temperature dependence of $C_\text{e}$ or $G$ in solid palladium for temperatures higher than 1500~K, as its melting point at 1 bar of pressure is 1828.1~K \cite{Lide2000}. Therefore, our first approach 
was to use the available models in the literature based on these low temperature ($<$1500 K) experimental data. In case of $G$, we neglected its $T_\text{e}$ dependence and used the constant value $G_{1}=8.95\times$10$^{17}$ WK$^{-1}$m$^{-3}$, which was obtained in Ref.~\onlinecite{Szymanski2007} from the phonon spectrum second moment measured at 296~K~\cite{Miiller1971}. In case of $C_\text{e}$, its  $T_\textrm{e}$-dependence was described by
\begin{equation}
\label{eq:ce}
 C_\mathrm{e,1}(T_\textrm{e})=T_\mathrm{e}\gamma(T_\mathrm{e})   
\end{equation}
where 
\begin{align}
    \label{eq:gamma}
    \gamma(T_\mathrm{e})=a_0& +a_1\exp\left(-\frac{T_\mathrm{e}-T_1}{\sigma_1}\right)
    +a_2\cosh^{-2}\left( \frac{T_\mathrm{e}-T_2}{\sigma_2} \right) \nonumber\\
    &+a_3\cosh^{-2}\left( \frac{T_\mathrm{e}-T_3}{\sigma_3} \right)
\end{align}
is the expression we employed in Ref.~\onlinecite{Muzas2024} to fit the available experimental data with the following parameters: $a_0=273$, $a_1=2702$, $a_2=214$, $a_3=98189$ (all in JK$^\mathrm{-2}$m$^\mathrm{-3}$); $T_1=-96$, $T_2=213$, $T_3=-2085$, $\sigma_1=50$, $\sigma_2=103$, $\sigma_3=634$ (all in K).

\begin{figure}
	\centering
	\includegraphics[width=1.0\columnwidth]{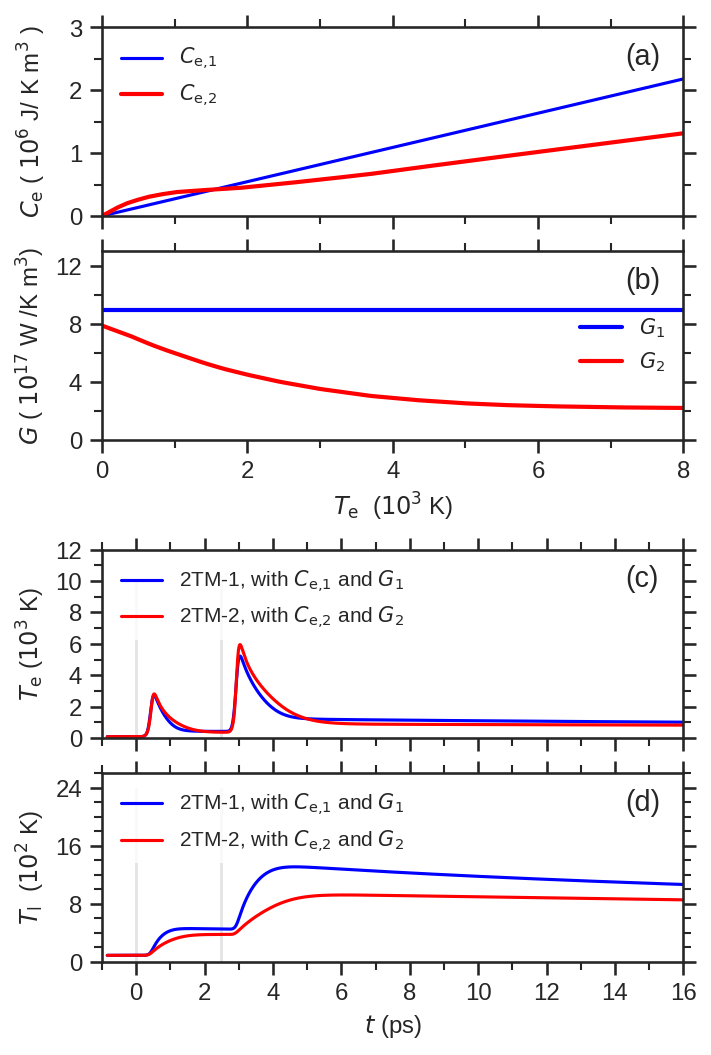}
    \caption{Electronic temperature dependence predicted by different models present in the literature for the palladium (a) electron heat capacity $C_\mathrm{e}$ and (b) electron-phonon energy exchange coupling factor $G$.  In blue, we show  $C_\text{e,1}$ and $G_1$ adjusted to experimental data for temperatures below 1500~K, extracted from Refs.~\onlinecite{Muzas2024} and \onlinecite{Szymanski2007}. In red, we show $C_\text{e,2}$ and $G_2$  extracted from DFT-based calculations in Ref.~\onlinecite{Li2022}. Illustrative 2TM temperature profiles obtained with said models in the case of two pulse irradiation with 38 and 93~J/m$^2$ of absorbed laser fluence and 2.5~ps of delay for (c) electronic temperature $T_\textrm{e}$ and (d) phononic temperature $T_\textrm{l}$.}
	\label{fig:CeG}
\end{figure}

The 2TM electronic temperature calculated using the above description of the macroscopic thermodynamical variables ($C_\textrm{e,1},C_\textrm{l},\kappa_\textrm{e},G_1$) allowed us to reproduce rather well the strong coverage dependence of the laser-induced CO photodesorption as well as the fluence dependence that were measured in single pulse experiments also reported by Hong \emph{et al}~\cite{Hong2016,Muzas2024}. 
However, first principles DFT-based calculations exploring the temperature dependence of $G$ and $C_\mathrm{e}$ in palladium~\cite{Li2022} within the temperature range from 300~K to 20\,000~K indicated that those quantities are being overestimated by about a factor 4 and 2, respectively, at the high electronic temperatures (6000$-$8000~K) that can be reached upon laser irradiation [see Figs.~\ref{fig:CeG}(a) and \ref{fig:CeG}(b)]. The general impact of using these first principles thermodynamical variables (hereinafter called $C_\textrm{e,2}$ and $G_2$, instead of  $C_\textrm{e,1}$ and $G_1$) in the calculated 2TM temperatures is an increase of the maximum value of $T_\text{e}$  and a concomitant decrease of $T_\text{l}$ [see Figs.~\ref{fig:CeG} (c) and ~\ref{fig:CeG}(d)] that is heavily affected by the laser fluence and time delay between pulses. Whether these differences affect or not the results of 2PC simulations will be discussed in Sec.~\ref{sec:results}. 
In order to avoid ambiguities, hereinafter, 2TM calculations performed with $C_\text{e,1}$ and $G_1$ parameterizations will be denoted as 2TM-1, whereas 2TM calculations performed with $C_\text{e,2}$ and $G_2$ will be denoted as 2TM-2. 

\subsection{\label{sec:eta_Te_model} Dependence of the friction coefficient on T$_\textrm{e}$}

Most of our analysis is based on adsorbate electronic friction coefficients that are independent of $T_\textrm{e}$, which is usually a reasonable approximation for metal substrates. Nevertheless, we will also investigate how the results of the simulations are affected by considering the dependence on $T_\textrm{e}$ of the strength of the adsorbate-electron coupling. With this aim, 
we consider explicitly the $T_\textrm{e}$ dependence of the adsorbate friction coefficient in Eq.~\eqref{eq:langevin}, consistently with inclusion of temperature-dependent $G(T_\mathrm{e})$ and $C_\mathrm{e}(T_\mathrm{e})$ in 2TM-2. Following Ref.~\onlinecite{Hellsing1984}, the mode-resolved friction tensor component $\eta_{\lambda}$ reads
\begin{equation}
    \eta_\lambda = \sum_\mathbf{q}^{N_\mathrm{q}} w_\mathbf{q}\eta_{\lambda,\mathbf{q}} = 
    \lim_{\omega_{\lambda,\mathbf{q}}\to 0} \sum_\mathbf{q}^{N_\mathrm{q}} w_\mathbf{q}
    \frac{M_{\lambda,\textbf{q}}}{\hbar}
    \Gamma_{\lambda}(\mathbf{q},\omega_{\lambda,\textbf{q}})
    \label{eq:eta-width}
\end{equation}
with $\lambda$ the phonon mode index; $\textbf{q}$ and $\omega_{\lambda,\textbf{q}}$ the phonon momentum and energy, respectively; $M_{\lambda,\mathbf{q}}$ and $\Gamma_{\lambda}(\mathbf{q},\omega_{\lambda,\textbf{q}})$ the mass and linewidth of the ($\lambda$,$\mathbf{q}$)-mode, respectively; and $N_\mathbf{q}$  the number of $\mathbf{q}$-points in the phonon Brillouin zone with weight $w_\mathbf{q}$. The linewidth is obtained from the imaginary part of the phonon self-energy calculated up to first-order in the electron-phonon coupling, $\pi_\lambda(\mathbf{q},\omega_{\lambda,\textbf{q}})$, as
\begin{equation}
   \Gamma_{\lambda}(\mathbf{q},\omega_{\lambda,\textbf{q}})= 2\hbar \mathrm{Im} \pi_\lambda(\mathbf{q},\omega_{\lambda,\textbf{q}}),
   \label{eq:width}
\end{equation}
where the mode resolved phonon self-energy reads
\begin{align}
    \pi_{\lambda} (\mathbf{q},\omega_{\lambda,\textbf{q}})= 
    \sum_{\mathbf{k}=0}^{N_\mathbf{k}}\sum_{\substack{\mu,\nu=1}}^{N_\textrm{b}}
  &  w_\mathbf{k}\left|g_\lambda^{\mu,\nu}(\textbf{k},\textbf{q})\right|^2 \nonumber \\
  &  \times  \frac{f(\epsilon_{\mu,\textbf{k}})-f(\epsilon_{\nu,\textbf{k+q}})}
    {\epsilon_{\nu,\textbf{k+q}}-\epsilon_{\nu,\textbf{k+q}}-\omega_{\lambda,q}-i\delta} 
\end{align}
with ($\mu$, $\nu$), $\textbf{k}$, and $\epsilon$ are electron band index, momentum, and energy, respectively, 
$N_\textrm{b}$ is the total number of electronic bands considered, and 
$N_\mathbf{k}$ is the number of $\mathbf{k}$-points in the electronic Brillouin zone with weight $w_\mathbf{k}$.
The function $f(\epsilon_{\mu\textbf{k}})=1/(e^{\beta(\epsilon_{\mu\textbf{k}}-\mu(T_\textrm{e}))}+1)$ is the Fermi-Dirac distribution function, where $\beta=1/(k_BT_\textrm{e})$, $k_B$ is the Boltzmann constant, $T_\textrm{e}$ is the electronic temperature, and $\mu(T_\textrm{e})$ is the chemical potential at $T_\textrm{e}$.  
Similarly to previous works~\cite{Bombin2023,Bombin2023L}, we fix the broadening parameter to a finite, physically motivated value of 30~meV~\cite{Hayashi2013,Schendel2017}. 
Note that summation over spin-degree of freedom is omitted for simplicity. 

The friction tensor component corresponding to the $j$th cartesian atomic coordinate, 
$\mathbf{e}_{j,\mathbf{q}}$=$(0_1,...,0_{j-1},1_j,0_{j+1},...,0_{3N})_\mathbf{q}$, with $N$
the number of atoms in the system reads
\begin{align}
\eta_{j} 
& =\sum_\mathbf{q}^{N_\mathrm{q}} w_\mathbf{q} \eta_{j\mathbf{q}} 
= \sum_\mathbf{q}^{N_\mathrm{q}} w_\mathbf{q}\bra{\mathbf{e}_{j,\mathbf{q}}} \hat{\eta} \ket{\mathbf{e}_{j,\mathbf{q}}} \nonumber \\
& = \sum_\mathbf{q}^{N_\mathrm{q}} w_\mathbf{q}\sum_\lambda^{N_{\mathrm{modes}}}  \left|v^j_{\lambda,\mathbf{q}}\right|^2
\bra{\lambda\mathbf{q}} \hat{\eta} \ket{\lambda\mathbf{q}}
    \label{fric:tensoratom}
\end{align}
where using Dirac notation  
$\bra{\mathbf{e}_{j\mathbf{q}}} \hat{\eta} \ket{\mathbf{e}_{j\mathbf{q}}}$ and $\bra{\lambda\mathbf{q}} \hat{\eta} \ket{\lambda\mathbf{q}}$
are $\eta_{j\mathbf{q}}$ and $\eta_{\lambda\mathbf{q}}$, respectively, and 
$v^j_{\lambda,\mathbf{q}}= \braket{\mathbf{e}_{j,\mathbf{q}}|\lambda\mathbf{q}}$ the $j$th cartesian component of the $(\lambda, \textbf{q})$ normal mode. Note that although the cartesian basis ${\ket{\mathbf{e}_{j,\mathbf{q}}}}$ are equal for each $\textbf{q}$, we maintain the $\mathbf{q}$-index for analogy with the matrix elements in the normal mode basis and to remark  that the projection is done for the $q$-modes on each case. 
Also note that in our implementation electron-phonon coupling is considered up to first order only, with no direct coupling between normal modes, and thus, the off-diagonal components of the friction tensor are zero in the normal mode basis ($\bra{\lambda\mathbf{q}} \hat{\eta} \ket{\lambda^\prime\mathbf{q}^\prime} = 0$ for $\lambda\ne\lambda^\prime$ or $\mathbf{q}\ne\mathbf{q}^\prime$)~\cite{Novko2016, Novko2018}.  In our MDEF simulations, the $T_\textrm{e}$ dependence is included in the LDFA friction coefficient for each atom as the average of the three diagonal components of the friction tensor corresponding to that atom, i.e., $\eta_\mathrm{LDFA}(T_\textrm{e})/\eta_\mathrm{LDFA}(0)=\langle \eta_j(T_\textrm{e})/\eta_j(0)\rangle$. 
Previous studies have evaluated the friction tensor following Eq.~\eqref{eq:eta-width} including off-diagonal components in the friction tensor to perform molecular dynamics simulations and show a small effect on the energy relaxation process~\cite{Askerka2016PRL,Askerka2017PRLerratum,Askerka2016PRB, Spiering2018}.

Evaluation of Eqs.~\eqref{eq:eta-width} to \eqref{fric:tensoratom} requires the knowledge of the electron band-structure, phonon dispersion, and electron-phonon matrix elements of the system. 
In our previous work, all these quantities were evaluated for Pd(111) with a CO coverage of 0.5~ML and $c(4\times2)$-2CO arrangement, considering two different adsorption structures in which the CO molecules were either adsorbed on bridge sites or on three-fold hollow positions~\cite{Bombin2023L,Bombin2023}. Considering the huge computational cost of these calculations, here we take advantage of that previous work and evaluate the $T_\textrm{e}$ dependence of $\eta$ for those two adsorption configurations, for which we already had calculated the electron and phonon structures and the electron-phonon matrix elements. In particular, 
we used the Quantum Espresso (QE) package~\cite{Giannozzi2009,Giannozzi2017} to evaluate the electronic band structure up to 6~eV above the Fermi level, enough to characterize thermal states at the high electronic temperatures generated in 2PC experiments. 
The phonon dispersion and perturbation potentials needed to evaluate the electron-phonon coupling matrix elements were evaluated using the \textsc{PHonon} code of the QE package at density functional perturbation theory level. Finally, we made use of the electron phonon Wannier (EPW) code to evaluate the electron-phonon matrix elements~\cite{NOFFSINGER20102140,PONCE2016116}. For further computational details we refer the reader to Ref.~\onlinecite{Bombin2023L}. 

\begin{figure}[h]
\includegraphics[width=1.0\linewidth]{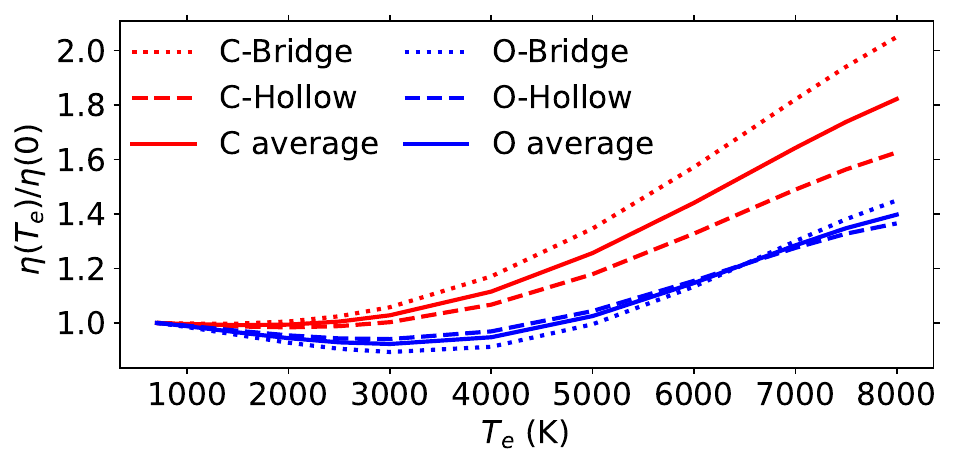}
\caption{\label{fig:etaTe} $T_\textrm{e}$-dependence of the atom-resolved electronic friction coefficient for CO on Pd(111). Dotted (dashed) lines corresponds to calculations where the CO is adsorbed at bridge (three-fold hollow) positions with a coverage of 0.5~ML. Their average is shown in solid lines. }
\end{figure}

Figure~\ref{fig:etaTe} shows the calculated $T_\textrm{e}$ dependence of the friction coefficient for C and O atoms in CO molecules adsorbed at three-fold hollow and bridge sites on Pd(111). 
For temperatures below $T_\textrm{e}= 4000$~K, the friction coefficients remain essentially constant. However, as $T_\textrm{e}$ increases to values reached in 2PC experiments, they also increase. 
At the highest temperature considered ($T_\textrm{e}= 8000$~K), the average increase in the friction coefficient (solid lines in the figure) is a factor 1.4 for O and 1.8 for C. In our LDFA implementation, we extrapolate that average temperature dependence to other configurations. To do so $\eta(T_{\rm e})$ is fitted with a function of the form:
\begin{equation}
\frac{\eta(T_\textrm{e})}{\eta(0)} = 1 + b_2T_\textrm{e}^2 + b_3T_\textrm{e}^3 + b_4T_\textrm{e}^4
\label{eq:fiteta}
\end{equation}
where the temperature is expressed in Kelvin. The corresponding fitting parameters ($b_i$, $i = 2,3,4$) for C and O are given in Table~\ref{tab:fiteta}.
\begin{table}
    \centering
    \begin{ruledtabular}
    \begin{tabular}{cccccc}
          & $b_1$                & $b_2$  & $b_3$   & $b_4$ \\
          & $\times$10$^{-4}$ K$^{-1}$ &    $\times$10$^{-8}$K$^{-2}$ & $\times$10$^{-12}$K$^{-3}$ & $\times$10$^{-16}$K$^{-4}$\\ \hline
        C &   $-$                       &-1.7301 &   8.3003 & -5.6924\\
        O &   $-$                       &-3.5923 &  10.679  & -6.8137\\
        Pd&  -2.8080                  & 3.7974 & -1.9487  &  0.22991 
    \end{tabular}
    \caption{Fitting parameters for the $T_\mathrm{e}$-dependence of the C and O friction coefficients $\eta(T_{\rm e})$ [Eq.~\eqref{eq:fiteta}] and the $T_\mathrm{e}$-dependence of the Pd electron-phonon coupling factor $G(T_{\rm e})$ [Eq.~\eqref{eq:G_Te}].
    }
    \label{tab:fiteta}
    \end{ruledtabular}
\end{table}

\subsection{\label{sec:system} Modeling the covered Pd(111) surface}

In the MD simulations the CO-covered Pd(111) surface is modeled with four Pd layers and the CO adlayer adsorbed on the topmost layer. Three of those layers as well as the adlayer are allowed to move following Eq.~\eqref{eq:langevin}, whilst the Pd bottom layer is kept frozen. We use 4$\times$2 supercells for the 0.75~ML coverage and 3$\times$3 and 6$\times$6 supercells for the MD simulations of the 0.33~ML coverage that are based on 2TM-1 and 2TM-2, respectively. In comparing the results for the low coverage we took into account that the use of different cell sizes affects the value of the desorption probability (roughly by a factor 2), but not its dependence on the absorbed laser fluence~\cite{Muzas2024}. Details of the adlayer structures can be found elsewhere~\cite{Hong2016, alducin2019,Muzas2024}, hence they are briefly summarized here. Starting with the saturation coverage (0.75~ML), the (vertically) adsorbed CO molecules are equally distributed between top, hcp, and fcc sites, with the strength of the binding energy on each site increasing in this order. In the 0.33~ML, all CO molecules are strongly bound on fcc sites.
%
\section{\label{sec:results} Two-pulse correlation simulations: CO/P\lowercase{d}(111)}

Motivated by the 2PC measurements reported by Hong \textit{et al.}~\cite{Hong2016}, we have carried out $(T_\mathrm{e},T_\mathrm{l})$-MDEF simulations using different time-delays $\delta t$ that vary between $-20$ and 20~ps for two surface coverages, the experimental saturation coverage of 0.75~ML and a low coverage of 0.33~ML that will be compared to the experimental results for 0.24~ML. Since the surface coverage of 0.24~ML consists of patches of 0.33~ML we still expect a qualitative agreement between our simulations and experiments, but not a quantitative agreement because the CO adsorbates are slightly less bound in the patches than in the pure 0.33~ML. This was what we obtained for instance when comparing the theoretical and experimental results regarding the fluence dependence of the CO photodesorption  induced by single-pulses~\cite{Muzas2024}. 

The features of the strong and weak laser pulses follow the experimental ones~\cite{Hong2016}. In particular, both are pulses of 780~nm, FWHM$=130$~fs, and $\mathrm{sech^{2}}$ temporal profile, with the absorbed fluences being 93 and 38~J/m$^2$ for 0.33~ML and 51 and 21~J/m$^2$ for 0.75~ML. Note that at zero delay the total laser fluences are 131~J/m$^2$ for the low coverage and 72~J/m$^2$ for the saturation coverage. As in experiments, the covered surfaces are initially equilibrated at 90~K. For each coverage and time-delay, the CO desorption probability is calculated from a set of 1000--5000 trajectories that are integrated up to 50--100~ps using the Beeman algorithm and a time integration step of 0.2~fs. In each case, the employed integration conditions assure that the calculated values are converged in time and statistically meaningful. 
We start our analysis by comparing the results obtained with the 2TM-1 and 2TM-2 parameterizations described in Sec.~\ref{sec:system}. The MDEF-1 results corresponds to  $(T_\mathrm{e},T_\mathrm{l})$-MDEF simulations based on the $T_\text{e}(t)$ values obtained from 2TM-1, in which $G(T_\textrm{e})$ is assumed to be constant and $C_\textrm{e}(T_\textrm{e})$ is described by fitting Eqs.~\eqref{eq:ce} and \eqref{eq:gamma} to experimental data measured up to 1500~K. In  MDEF-2, the $(T_\mathrm{e},T_\mathrm{l})$-MDEF simulations are based on 2TM-2, in which we account for the $T_\textrm{e}$-dependence of $G(T_\textrm{e})$ and $C_\textrm{e}(T_\textrm{e})$ calculated from first-principles~\cite{Li2022}. To single out the sensitivity of the $(T_\mathrm{e},T_\mathrm{l})$-MDEF simulations to the 2TM parametrization, in these first sets of simulations the friction coefficients of the adsorbates are the LDFA values calculated at $T_\textrm{e}=0$~K [$\eta(0)$].

\begin{figure}[t!]
\includegraphics[width=1.0\columnwidth]{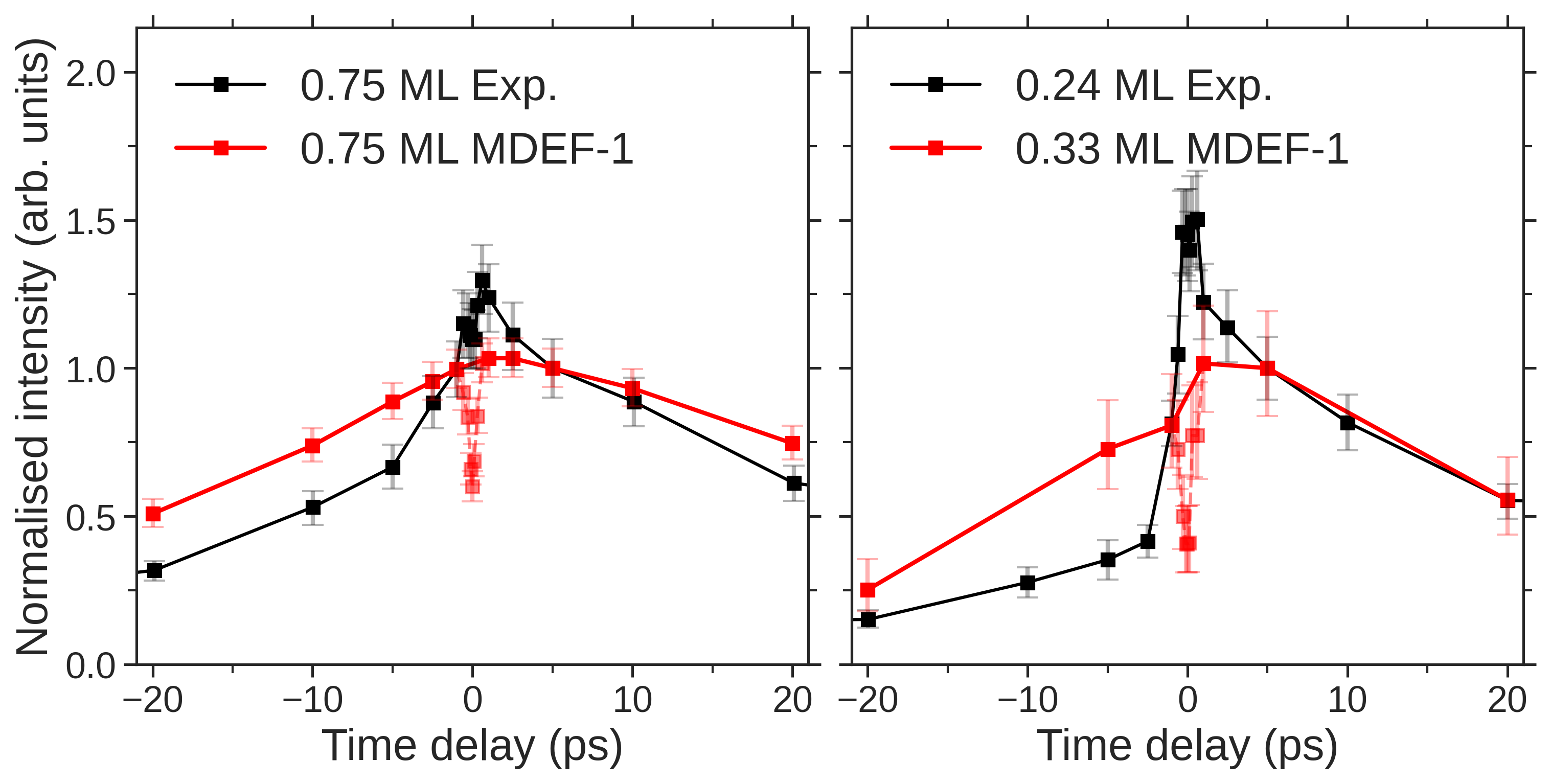}
\caption{\label{fig:2pc_camillone} CO desorption probability as a function of time delay $\delta t$ between the high and low intensity pulses obtained in our $(T_\mathrm{e},T_\mathrm{l})$-MDEF simulations based on 2TM-1 and $\eta(0)$ for the adsorbates friction coefficients (red squares and lines) and in two-pulse correlation measurements from Ref.~\onlinecite{Hong2016} (black squares and lines) normalized to $\delta t=5$~ps. Positive (negative) delays correspond to the low intensity pulse arriving first (second). Left: the initial coverage is 0.75~ML in both experiments and simulations. Right: the initial coverage is 0.24~ML in experiments and 0.33~ML in our simulations.}
\end{figure}
Figure~\ref{fig:2pc_camillone} shows the desorption probability as a function of time delay for 0.75~ML and 0.33~ML that we obtain from MDEF-1. 
As shown in Ref.~\onlinecite{Muzas2024}, the calculated desorption probabilities at 0.75~ML (0.33~ML) are about a factor 2 (6) smaller than in the experiments at 0.75~ML (0.24~ML). Thus, to facilitate the comparison with the 2PC measurements and focus on the qualitative behavior, each set of data is normalized to its corresponding value at 5~ps of time delay. As usual, the theoretical results close to zero delay largely underestimate the experimental values. This behavior is typically discussed in terms of possible limitations of 2TM at very large fluences and we will return to this point at the end of this section. Focusing on the comparison beyond the subpicosencond delay, the simulations capture the asymmetry between positive and negative delays and the fact that this asymmetry is more pronounced at the low coverage (0.33 ML in simulations and 0.24~ML in experiments). Still, the comparison is not satisfactory enough. Our simulations cannot reproduce the sharper decrease of the probabilities that is measured at negative delays at 0.75~ML and particularly at 0.24~ML. The results seem however reasonable in view of the differences of the electronic and lattice temperatures between positive and negative delays that are obtained from 2TM-1 (Figs.~S1 and S2 in the supplementary material). Then, as a next step we have repeated the simulations but using the electronic temperature calculated with the 2TM-2 parameterization, whose values between positive and negative delays are in fact more asymmetric (Figs.~S3 and S4 in supplementary material).

\begin{figure}
\includegraphics[width=1.0\columnwidth]{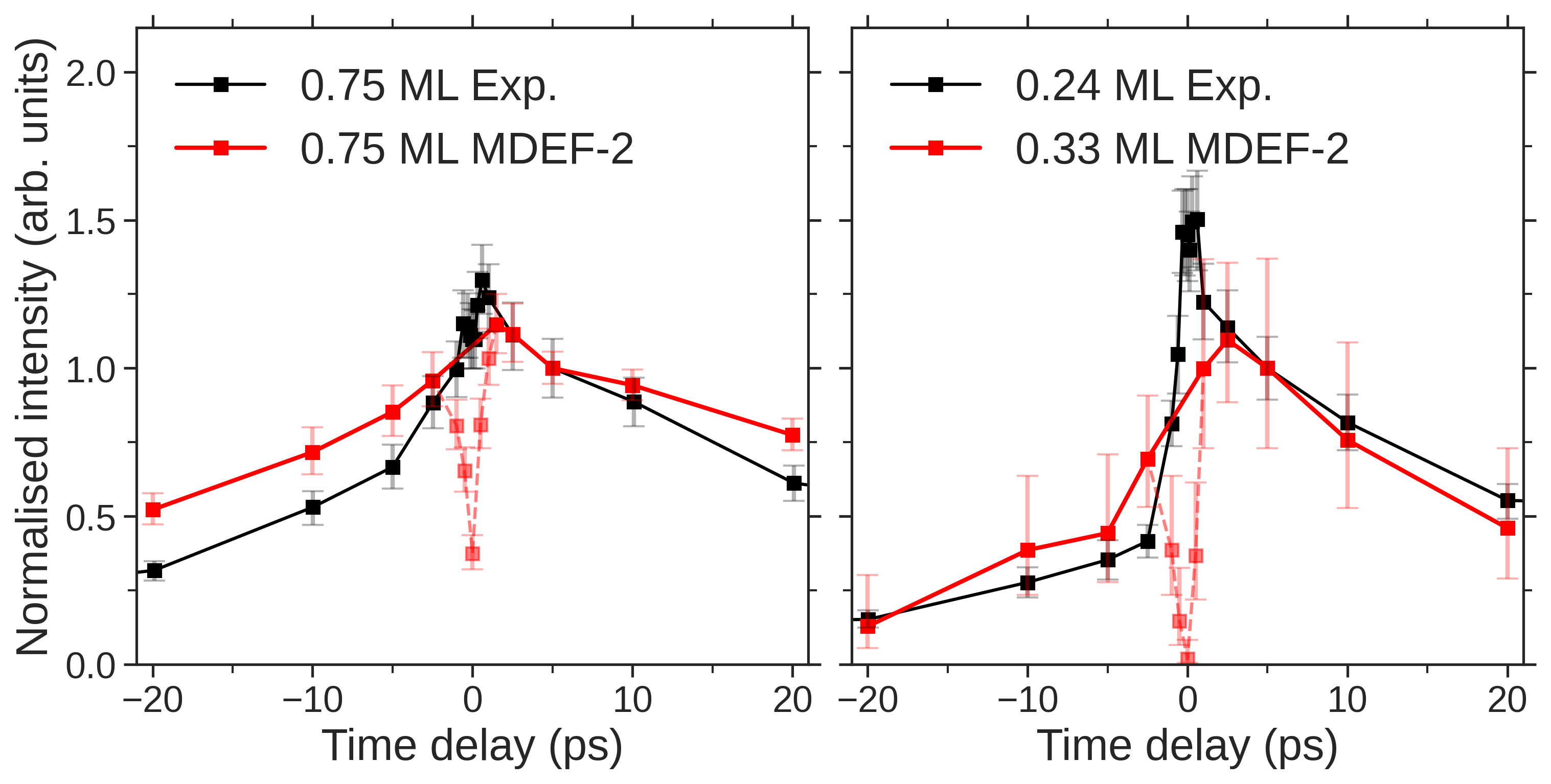}
\caption{\label{fig:2pc_liji} Same as Fig.~\ref{fig:2pc_camillone} but for our $(T_\mathrm{e},T_\mathrm{l})$-MDEF simulations based on the electronic temperature calculated with 2TM-2 and the friction coefficients at 0~K for C and O ($\eta(0)$).}
\end{figure}
The new results obtained from MDEF-2 simulations 
are shown in Fig.~\ref{fig:2pc_liji}. As in the previous figure, each set of data is normalized to its corresponding value at 5~ps of time delay to facilitate the comparison with the 2PC measurements. Now the (qualitative) agreement between theory and experiments at delays greater than 1~ps is somewhat better. In general, the MDEF-2 results exhibit a slightly larger asymmetry between positive and negative delays than those of Fig.~\ref{fig:2pc_camillone}. For 0.75~ML, the new simulations reproduce rather well the increase in the desorption probability from 5~ps to 1.5~ps delay that is observed in experiments. For 0.33~ML, the MDEF-1 and MDEF-2 results are qualitatively similar at positive delays, but the difference between positive and negative delays in MDEF-2 is in better agreement with the high asymmetry measured for 0.24~ML. Next, considering that these latter simulations improve the comparison to experiments, we analyze in more detail the MDEF-2 results to understand the asymmetry in the 2PC measurements and its dependence on the CO coverage. 

\begin{figure}
\includegraphics[width=1.0\columnwidth]{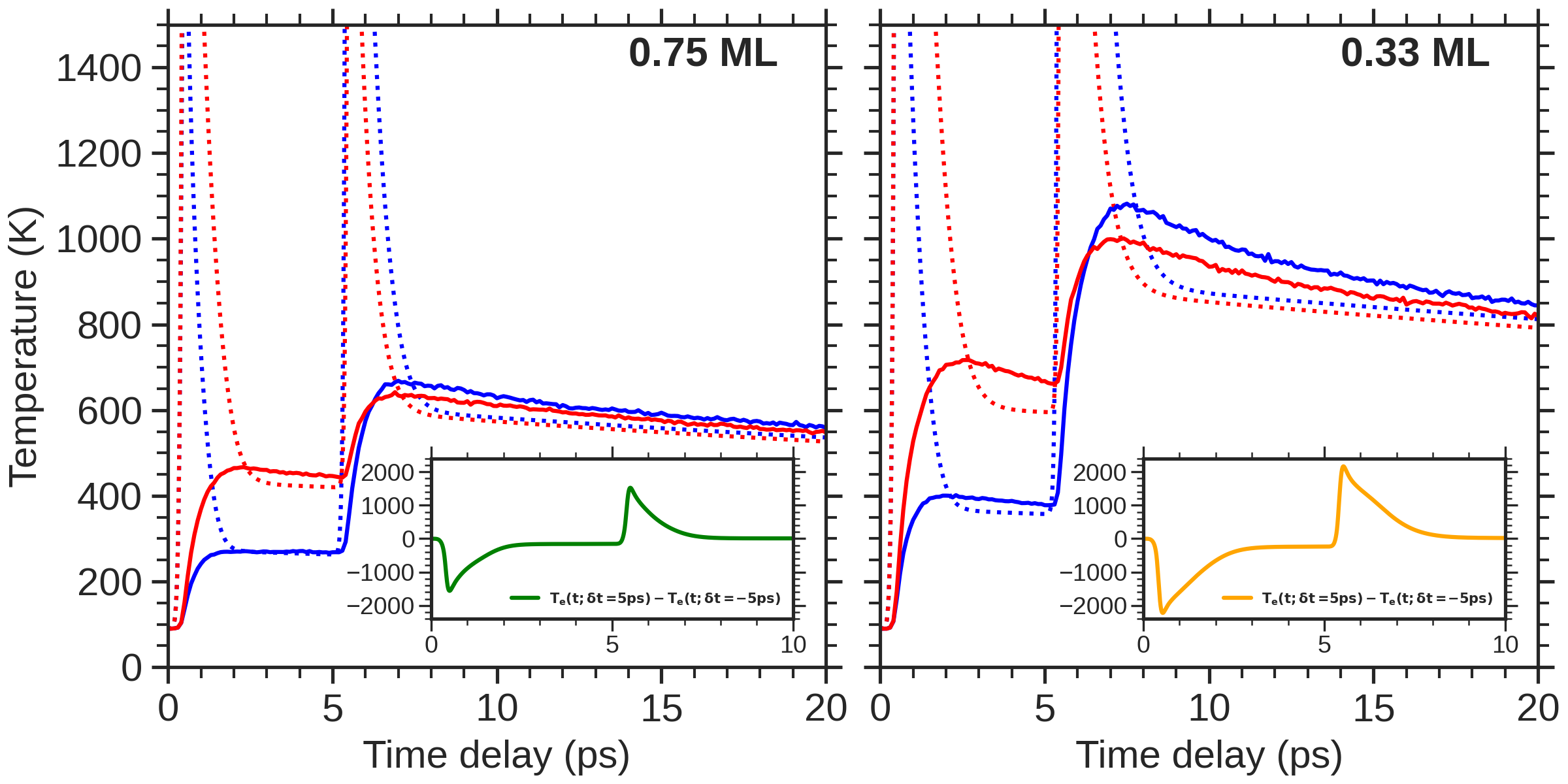}
\caption{\label{fig:t_ads} Positive-to-negative delay comparison of $T_\textrm{ads}(t)$ (solid lines) and  $T_\textrm{e}(t)$ (dashed lines) for 0.75~ML (left panel) and 0.33~ML (right panel). Results obtained with 2TM-2 and MDEF-2 for time delays $\delta t=\!5$~ps (blue lines) and  $\delta t=\!-5$~ps (red lines). Insets: Difference between $T_\textrm{e}(t)$ at $\delta t=\!5$ and $\delta t=\!-5$~ps as obtained for the laser pulses employed at 0.75~ML (green, left inset) and 0.33~ML (orange, right inset).}
\end{figure}

The adsorbate temperature $T_\textrm{ads}(t)$ plotted in Fig.~\ref{fig:t_ads} for each coverage and at time delays $\delta t\!=\!\pm 5$~ps provides direct evidence on how the adsorbates become differently excited at negative and positive delays. (The figure is representative of what we obtain at different delays, as depicted in more detail in Figs.~S5 and S6, supplementary material). At negative delays the first pulse is the most intense and causes a higher excitation in the adsorbates (higher $T_\textrm{ads}$) than the corresponding positive delay. However, this initial excitation seems insufficient to induce CO desorption even at negative delays, for which $T_\textrm{ads}$ is transiently as high as $\sim750$~K in the low coverage case. The analysis of the desorption probability as a function of time $P_\textrm{des}(t)$ shows that desorption is basically initiated 1-2 ps after arrival of the second pulse see Figs.~S7 and S8 in the supplementary material). And the latter occurs not only for positive, but also for negative delays, for which the first pulse is the most intense. Thus, this behavior of $P_{\rm des}(t)$ suggests that it is the electronic excitation created by the second pulse, and not the transient excitation created by the first pulse, that finally triggers the desorption process. Then, focusing on the excitation created by the second pulse in the adsorbates, Fig.~\ref{fig:t_ads} shows that this excitation is higher (higher $T_\textrm{ads}$) at positive delay, and thus the desorption probability (Fig.~\ref{fig:2pc_liji}). 

Regarding the dependence on CO coverage, the larger asymmetry observed at 0.33~ML is a consequence of various factors. Since the total laser fluence employed for the low coverage doubles that of the saturation coverage, the corresponding $T_\textrm{e}$ is considerably higher at these large fluences (1000--2000~K higher if we compare the maximum values as shown in Figs.~S3 and S4 in the supplementary material). Furthermore, the asymmetry in $T_\textrm{e}(t)$ is more pronounced at the large fluence (compare the insets in Fig.~\ref{fig:t_ads}). These two factors are common to both coverages, but the asymmetry they can cause in the 2PC measurements is particularly enhanced in the low coverage. The reason is that, as found in experiments~\cite{Hong2016} and reproduced by our single-pulse simulations~\cite{Muzas2024},  the low coverage is characterized by the strongest dependence of the desorption probability on the laser fluence. Therefore, a similar change in $T_\textrm{e}$ causes larger differences in this coverage. 

As mentioned above, although our $(T_\mathrm{e},T_\mathrm{l})$-MDEF simulations based on 2TM-2 reproduce reasonably well the asymmetric 2PC measurements of the desorption probability in an extensive range of time delays, they are not able to capture the subpicosecond behavior. In this case, the experiments show a rapid increase of the desorption yield, more apparent in the case of low coverage for which a narrow prominent peak appears at subpicosecond delays together with a small dip at zero delay, whereas the simulations show a strong reduction of the desorption probability in this range of delays. In evaluating this discrepancy, we have to consider the limitations of the theoretical model and also the experimental uncertainties that are already discussed by Hong et al~\cite{Hong2016}. In particular, the authors mentioned that near zero delay the residual
optical interference between the two pulses can generate hot spots
that might be artificially enhancing the reported desorption
probabilities. Having this in mind, we can still think in different reasons that might be responsible for a possible failure of the theoretical model at subpicosecond delays. One of these reasons is the fact that the employed adsorbate electronic friction coefficients are calculated at zero temperature. This approximation is often justified by the fact that the Fermi temperature of metals is very high. However, at zero time delay the achieved $T_\textrm{e}$ are also very high (around 6000~K in 0.33~ML and around 4000~K in 0.75~ML). As a result, the failure to capture the zero time delay peak may be related to the underestimation of the adsorbate-electron coupling at high electronic temperatures. A stronger adsorbate electron-coupling would make the adsorbates respond more rapidly to the high electronic temperatures, which would be expected to lead to an increased desorption yield at zero time delay. Another possible reason is the assumption made by 2TM that the electronic excitation can be described by an electronic temperature. As discussed above, this assumption is well justified due to the rapid thermalization of electronic excitations in metals due to the efficient electron-electron scattering. However, for very short time delays the initial non-thermal electronic excitation generated by the laser pulses may still bring about a measurable effect.
\begin{figure}
\includegraphics[width=1.0\linewidth]{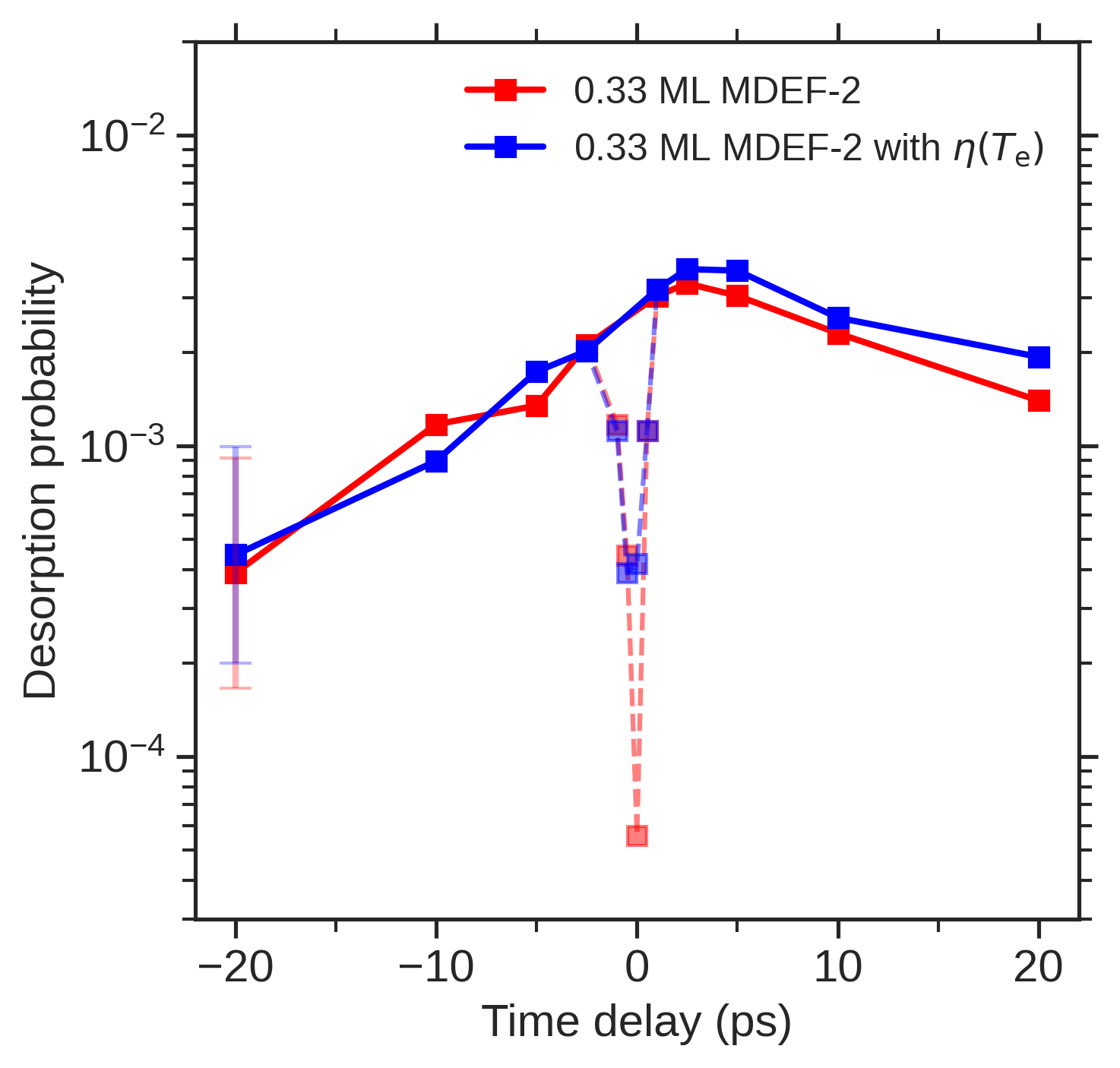}
\caption{\label{fig:2pc_etaTe} CO desorption probability
as a function of time delay for the 0.33~ML coverage. Results obtained from MDEF-2 simulations using for C and O the friction coefficients at 0~K (red curve) and the $T_\textrm{e}$-dependent friction coefficients (blue curve).}
\end{figure}

In the following, we analyze whether the first of these considerations, the neglect of the increase of the strength of the adsorbate-electron coupling with increasing temperature, is the reason why the simulations do not capture the large desorption yield increase at subpicosecond time delays. With this aim, we have performed simulations using the 2TM-2 model with the temperature-dependent electronic friction coefficients $\eta(T_\textrm{e})$ described and calculated for CO in Sec.~\ref{sec:eta_Te_model}. Figure~\ref{fig:etaTe} shows that it is only at $T_\textrm{e} \ge 5000$~K that $\eta(T_\textrm{e})$ starts to differ in more than a factor 1.2 from its value at 0~K. Therefore, these new simulations were only done for the low coverage, for which $T_\textrm{e}$ can reach high values of about 5000--6000~K at certain time delays. For 0.75~ML, we do not expect significant changes because the highest $T_\textrm{e}$ values are around 4000~K at most. In Fig.~\ref{fig:2pc_etaTe} we show the comparison of the results obtained for the 0.33~ML case with and without $T_\textrm{e}$-dependent friction coefficients. As expected, the results are very similar at suprapicosecond time delays, for which the agreement with the experiment was obtained. At subpicosecond time delays, a noticeable increase in the desorption yield that amounts around an order of magnitude at zero delay is obtained when the $T_\textrm{e}$-dependent friction coefficients are employed. This shows that the underestimation of the strength of the adsorbate-electron coupling at high temperatures was in part responsible of the disagreement between the simulations and experiments. 

However, the new simulations are still unable to capture the peak observed at the subpicosecond regime. On the contrary, they still predict a decrease of the desorption yield at zero delay. In light of our results, we can exclude that this disagreement is due to the underestimation of the increase of the adsorbate-electron coupling when increasing temperature. 
As already noted above, one of the reasons for the remaining disagreement could be the optical interference between the two pulses near zero delay that was already pointed out as a possibility by the experimentalists. Nevertheless, it could also be due to some of the approximations of the model. For instance, it may be related to the incomplete description of the electronic structure at the extreme electronic temperatures that are reached at subpicosecond delays. This can be specially important for a metal like Pd for which, since the chemical potential at $T_\textrm{e}=0$~K lies at the edge of the d-bands, its position varies strongly with the temperature. As a consequence, it could be expected a strong perturbation of the electronic structure at very high values of $T_\textrm{e}$ that may affect the strength of the interaction between the adsorbates and the substrate atoms. As mentioned above, another reason could be the inaccuracy of the description of the laser excitation provided by the 2TM at very short times. More precisely, the basic assumption of the 2TM is that the electronic excitation can be described by an electronic temperature and, therefore, it neglects the nonthermal transient electronic excitation that lasts for a very short time after the laser excitation~\cite{Kratzer2022,Lisowski2004}. It cannot be excluded that at subpicosecond time delays this nonthermal distribution plays an important role that is not described by 2TM. A theoretical description of the nonthermal distribution of the electrons generated by the laser and of their coupling to the adsorbates would be necessary to explore this possibility.

\section{\label{sec:summary}Summary}

We have performed Langevin molecular dynamics simulations [$(T_\mathrm{e},T_\mathrm{l})$-MDEF] using an accurate DFT-based machine learning potential energy surface 
to study the photodesorption of CO from Pd(111) and its dependence on surface coverage that was measured in two-pulse correlation (2PC) experiments. 
To describe the extreme non-equilibrium conditions induced by the intense laser pulses, we make use of a two-temperature model (2TM), which accounts for the electronic excitation created by the laser pulses and the dynamical energy balance between the electron and phonon baths. In solving the 2TM equations, we have used two different parameterizations. One of them includes improved descriptions of the temperature-dependent electronic heat capacity $C_e(T_\textrm{e})$ and electron-phonon coupling constant $G(T_\textrm{e})$, which are based on \textit{ab-initio} calculations for $T_\textrm{e}$ in the range of $300-20\,000$~K~\cite{Li2022}. 
The results of our MDEF simulations, which were run for each of the electronic temperatures obtained with these two different 2TM parameterizations, reveal that including the \textit{ab-initio} based $C_e(T_\textrm{e})$ and $G(T_\textrm{e})$ is key to reproduce the asymmetry in the desorption probability between positive and negative time delays that is measured experimentally. 

Still, our simulations suffers the well known problem in the theoretical treatment of 2PC experiments that consists in the underestimation of the desorption probability around zero time delay. 
From the theoretical side, this discrepancy may arise due to the extreme electronic temperatures that are reached at near zero time delay. To offer some light into this point, we have also performed $(T_\mathrm{e},T_\mathrm{l})$-MDEF simulations in which the friction coefficient of the adsorbates also depends on the electronic temperature, what improves the description of the energy exchange between the hot electronic bath and the CO molecules. In these new simulations, the predicted desorption probability at zero delay increases by an order of magnitude, reducing the difference between the theoretical and experimental values. 
However, the discrepancy does not disappear. The simulations still predict a dip of the desorption yield in the subpicosecond region where the experiments show a peak. This allows us to conclude that even if the neglect of the increase of the strength of the adsorbate-electron coupling with increasing electronic temperature is in part responsible of the underestimation of the desorption yield at short delays, the observed discrepancy between theory and experiment cannot be explained uniquely by it.  

All in all, our results hint to the importance of improving the electronic structure description at the extreme electronic temperature experimental conditions, whose effect can be enhanced in the case of Pd due to the sensitivity of the chemical potential to changes in $T_\textrm{e}$~\cite{Li2022,Bombin2023L}. 
In this respect, active theoretical research is being developed in order to improve the temperature dependence of the parameters entering the 2TM, such as, the electronic heat conductivity~\cite{Medvedev2024} and electron-phonon coupling constant~\cite{Akhmetov2023,Smirnov2025}. Lines for future theoretical improvements include accounting for additional thermal effects in the system~\cite{Bevillon2014,Guo2018,Zhang2021} and for the coupling of the adsorbates to the out of equilibrium non-thermal electron distributions that are expected to appear in the subpicosecond regime after laser excitation~\cite{Kratzer2022,Lisowski2004}. Simulations based on time-dependent DFT (TDDFT) can be a good alternative method to tackle these limitations. Although being computationally more demanding, TDDFT offers the advantage of being able to simultaneously describe electronic structure thermal effects and the transient, non-thermal electronic distributions.

\section{Supplementary Material}
The supplementary material contains additional information extracted from the $(T_\mathrm{e},T_\mathrm{l})$-MDEF simulations.

\acknowledgments{
Financial support was provided by the
Spanish MCIN$/$AEI$/$10.13039$/$501100011033$/$, FEDER Una manera de hacer Europa (Grant No.~PID2022-140163NB-I00), Gobierno Vasco-UPV/EHU (Project No.~IT1569-22), and the Basque Government Education Departments’ IKUR program, also co-funded by the European NextGenerationEU action through the Spanish Plan de Recuperación, Transformación y Resiliencia (PRTR). Computer resources were provided by the Donostia International Physics Center (DIPC) Supercomputing Center. R.B. acknowledges funding from ADAGIO (Advanced Manufacturing Research Fellowship Programme in the Basque – New Aquitaine Region) from the European Union’s Horizon 2020 research and innovation programme under the Marie Sklodowska  Curie cofund Grant Agreement No. 101034379.
}

\bibliography{refs}

@PREAMBLE{
 "\providecommand{\noopsort}[1]{}" 
 # "\providecommand{\singleletter}[1]{#1}%" 
}

@article{alducinpss17,
title = "Non-adiabatic effects in elementary reaction processes at metal surfaces",
journal = "Prog. Surf. Sci.",
volume = "92",
number = "4",
pages = "317 - 340",
year = "2017",
issn = "0079-6816",
doi = "https://doi.org/10.1016/j.progsurf.2017.09.002",
url = "http://www.sciencedirect.com/science/article/pii/S0079681617300230",
author = "Alducin, M. and D\'iez Mui\~no, R. and Juaristi, J. I."
}

@Article{alducin2019,
  author    = {Alducin, Maite and Camillone, Nicholas and Hong, Sung-Young and Juaristi, J. I.},
  title     = {Electrons and Phonons Cooperate in the Laser-Induced Desorption of {CO} from {Pd(111)}},
  doi       = {10.1103/PhysRevLett.123.246802},
  issue     = {24},
  pages     = {246802},
  url       = {https://link.aps.org/doi/10.1103/PhysRevLett.123.246802},
  volume    = {123},
  journal   = {Phys. Rev. Lett.},
  month     = {Dec},
  numpages  = {6},
  publisher = {American Physical Society},
  year      = {2019},
}

@article{Anisimov1974,
author= { Anisimov, S. I. and Kapeliovich, B. L. and  Perel'Man, T. L. },
title={Electron emission from metal surfaces exposed to ultrashort laser pulses},
journal={Sov. Phys. JETP},
volume={39},
pages={375-377},
year={1974}
}

@book{Ashcroft1988,
  author = {Ashcroft, N. W. and Mermin, N. D.},
  publisher = {Holt-Saunders},
  title = {{S}olid {S}tate {P}hysics},
  year = 1988
}

@article{Askerka2016PRL,
  title = {Role of Tensorial Electronic Friction in Energy Transfer at Metal Surfaces},
  author = {Askerka, Mikhail and Maurer, Reinhard J. and Batista, Victor S. and Tully, John C.},
  journal = {Phys. Rev. Lett.},
  volume = {116},
  issue = {21},
  pages = {217601},
  numpages = {5},
  year = {2016},
  month = {May},
  publisher = {American Physical Society},
  doi = {10.1103/PhysRevLett.116.217601},
  url = {https://link.aps.org/doi/10.1103/PhysRevLett.116.217601}
}

@article{Askerka2017PRLerratum,
  title = {Erratum: Role of Tensorial Electronic Friction in Energy Transfer at Metal Surfaces [Phys. Rev. Lett. 116, 217601 (2016)]},
  author = {Askerka, Mikhail and Maurer, Reinhard J. and Batista, Victor S. and Tully, John C.},
  journal = {Phys. Rev. Lett.},
  volume = {119},
  issue = {6},
  pages = {069901},
  numpages = {2},
  year = {2017},
  month = {Aug},
  publisher = {American Physical Society},
  doi = {10.1103/PhysRevLett.119.069901},
  url = {https://link.aps.org/doi/10.1103/PhysRevLett.119.069901}
}

@article{Askerka2016PRB,
  title = {Ab initio tensorial electronic friction for molecules on metal surfaces: Nonadiabatic vibrational relaxation},
  author = {Maurer, Reinhard J. and Askerka, Mikhail and Batista, Victor S. and Tully, John C.},
  journal = {Phys. Rev. B},
  volume = {94},
  issue = {11},
  pages = {115432},
  numpages = {15},
  year = {2016},
  month = {Sep},
  publisher = {American Physical Society},
  doi = {10.1103/PhysRevB.94.115432},
  url = {https://link.aps.org/doi/10.1103/PhysRevB.94.115432}
}

@article{Bevillon2014,
  title = {Free-electron properties of metals under ultrafast laser-induced electron-phonon nonequilibrium: A first-principles study},
  author = {B\'evillon, E. and Colombier, J. P. and Recoules, V. and Stoian, R.},
  journal = {Phys. Rev. B},
  volume = {89},
  issue = {11},
  pages = {115117},
  numpages = {11},
  year = {2014},
  month = {Mar},
  publisher = {American Physical Society},
  doi = {10.1103/PhysRevB.89.115117},
  url = {https://link.aps.org/doi/10.1103/PhysRevB.89.115117}
}

@article{Bombin2023L,
  title = {Anomalous transient blueshift in the internal stretch mode of {CO/Pd(111)}},
  author = {Bomb\'{\i}n, Ra\'ul and Muzas, A. S. and Novko, Dino and Juaristi, J. I\~naki and Alducin, Maite},
  journal = {Phys. Rev. B},
  volume = {107},
  issue = {12},
  pages = {L121404},
  numpages = {6},
  year = {2023},
  month = {Mar},
  publisher = {American Physical Society},
  doi = {10.1103/PhysRevB.107.L121404},
  url = {https://link.aps.org/doi/10.1103/PhysRevB.107.L121404}
}

@article{Bombin2023,
  title = {Vibrational dynamics of {CO} on {Pd(111)} in and out of thermal equilibrium},
  author = {Bomb\'{\i}n, Ra\'ul and Muzas, A. S. and Novko, Dino and Juaristi, J. I\~naki and Alducin, Maite},
  journal = {Phys. Rev. B},
  volume = {108},
  issue = {4},
  pages = {045409},
  numpages = {14},
  year = {2023},
  month = {Jul},
  publisher = {American Physical Society},
  doi = {10.1103/PhysRevB.108.045409},
  url = {https://link.aps.org/doi/10.1103/PhysRevB.108.045409}
}

@article{Bonn99,
author = {M. Bonn  and S. Funk  and Ch. Hess  and D. N. Denzler  and C. Stampfl  and M. Scheffler  and M. Wolf  and G. Ertl },
title = {Phonon- Versus Electron-Mediated Desorption and Oxidation of {CO} on {Ru(0001)}},
journal = {Science},
volume = {285},
number = {5430},
pages = {1042-1045},
year = {1999},
doi = {10.1126/science.285.5430.1042},
url = {https://www.science.org/doi/abs/10.1126/science.285.5430.1042}
}

@article{Budde91,
  title = {Femtosecond time-resolved measurement of desorption},
  author = {Budde, F. and Heinz, T. F. and Loy, M. M. T. and Misewich, J. A. and de Rougemont, F. and Zacharias, H.},
  journal = {Phys. Rev. Lett.},
  volume = {66},
  issue = {23},
  pages = {3024--3027},
  numpages = {0},
  year = {1991},
  month = {Jun},
  publisher = {American Physical Society},
  doi = {10.1103/PhysRevLett.66.3024},
  url = {https://link.aps.org/doi/10.1103/PhysRevLett.66.3024}
}

@article{caruso22,
author = {Fabio Caruso and Dino Novko},
title = {Ultrafast dynamics of electrons and phonons: from the two-temperature model to the time-dependent {B}oltzmann equation},
journal = {Advances in Physics: X},
volume = {7},
number = {1},
pages = {2095925},
year  = {2022},
publisher = {Taylor & Francis},
doi = {10.1080/23746149.2022.2095925},
url = {https://doi.org/10.1080/23746149.2022.2095925}
}

@article{Corkum1988,
  title = {Thermal Response of Metals to Ultrashort-Pulse Laser Excitation},
  author = {Corkum, P. B. and Brunel, F. and Sherman, N. K. and Srinivasan-Rao, T.},
  journal = {Phys. Rev. Lett.},
  volume = {61},
  issue = {25},
  pages = {2886--2889},
  numpages = {0},
  year = {1988},
  month = {Dec},
  publisher = {American Physical Society},
  doi = {10.1103/PhysRevLett.61.2886},
  url = {https://link.aps.org/doi/10.1103/PhysRevLett.61.2886}
}

@article{Deliwala95,
title = {Surface femtochemistry of {O$_2$} and {CO} on {Pt(111)}},
journal = {Chemical Physics Letters},
volume = {242},
number = {6},
pages = {617-622},
year = {1995},
issn = {0009-2614},
doi = {https://doi.org/10.1016/0009-2614(95)00791-2},
url = {https://www.sciencedirect.com/science/article/pii/0009261495007912},
author = {S. Deliwala and R.J. Finlay and J.R. Goldman and T.H. Her and W.D. Mieher and E. Mazur}
}

@article{Denzler2003,
  title = {Electronic Excitation and Dynamic Promotion of a Surface Reaction},
  author = {Denzler, D. N. and Frischkorn, C. and Hess, C. and Wolf, M. and Ertl, G.},
  journal = {Phys. Rev. Lett.},
  volume = {91},
  issue = {22},
  pages = {226102},
  numpages = {4},
  year = {2003},
  month = {Nov},
  publisher = {American Physical Society},
  doi = {10.1103/PhysRevLett.91.226102},
  url = {https://link.aps.org/doi/10.1103/PhysRevLett.91.226102}
}

@article{Dion2004,
  title = {Van der Waals Density Functional for General Geometries},
  author = {Dion, M. and Rydberg, H. and Schr\"oder, E. and Langreth, D. C. and Lundqvist, B. I.},
  journal = {Phys. Rev. Lett.},
  volume = {92},
  issue = {24},
  pages = {246401},
  numpages = {4},
  year = {2004},
  month = {Jun},
  publisher = {American Physical Society},
  doi = {10.1103/PhysRevLett.92.246401},
  url = {https://link.aps.org/doi/10.1103/PhysRevLett.92.246401}
}

@article{frischkorncr06,
author = {Frischkorn, Christian and Wolf, Martin},
title = {Femtochemistry at Metal Surfaces: Nonadiabatic Reaction Dynamics},
journal = {Chem. Rev.},
volume = {106},
number = {10},
pages = {4207-4233},
year = {2006},
doi = {10.1021/cr050161r},
url = {http://dx.doi.org/10.1021/cr050161r}
}

@article{Fuchsel2011,
author ="Füchsel, Gernot and Klamroth, Tillmann and Monturet, Serge and Saalfrank, Peter",
title  ="Dissipative dynamics within the electronic friction approach: the femtosecond laser desorption of {H2/D2} from {Ru(0001)}",
journal  ="Phys. Chem. Chem. Phys.",
year  ="2011",
volume  ="13",
issue  ="19",
pages  ="8659-8670",
publisher  ="The Royal Society of Chemistry",
doi  ="10.1039/C0CP02086A",
url  ="http://dx.doi.org/10.1039/C0CP02086A"
}

@article{Funk2000,
    author = {Funk, S. and Bonn, M. and Denzler, D. N. and Hess, Ch. and Wolf, M. and Ertl, G.},
    title = {Desorption of {CO} from{ Ru(001) }induced by near-infrared femtosecond laser pulses},
    journal = {J. Chem. Phys.},
    volume = {112},
    number = {22},
    pages = {9888-9897},
    year = {2000},
    month = {06},
    issn = {0021-9606},
    doi = {10.1063/1.481626},
    url = {https://doi.org/10.1063/1.481626}
}

@article{Giannozzi2009,
author = {Giannozzi, Paolo and Baroni, Stefano and Bonini, Nicola and Calandra, Matteo and Car, Roberto and Cavazzoni, Carlo and Ceresoli, Davide and Chiarotti, Guido L and Cococcioni, Matteo and Dabo, Ismaila and Dal Corso, Andrea and De Gironcoli, Stefano and Fabris, Stefano and Fratesi, Guido and Gebauer, Ralph and Gerstmann, Uwe and Gougoussis, Christos and Kokalj, Anton and Lazzeri, Michele and Martin-Samos, Layla and Marzari, Nicola and Mauri, Francesco and Mazzarello, Riccardo and Paolini, Stefano and Pasquarello, Alfredo and Paulatto, Lorenzo and Sbraccia, Carlo and Scandolo, Sandro and Sclauzero, Gabriele and Seitsonen, Ari P and Smogunov, Alexander and Umari, Paolo and Wentzcovitch, Renata M},
doi = {10.1088/0953-8984/21/39/395502},
issn = {09538984},
journal = {J. Phys. Condens. Matter},
month = {sep},
number = {39},
pages = {395502},
pmid = {21832390},
title = {{QUANTUM ESPRESSO: A modular and open-source software project for quantum simulations of materials}},
url = {https://iopscience.iop.org/article/10.1088/0953-8984/21/39/395502},
volume = {21},
year = {2009}
}

@article{Giannozzi2017,
	author = {Giannozzi, P and Andreussi, O and Brumme, T and Bunau, O and {Buongiorno Nardelli}, M and Calandra, M and Car, R and Cavazzoni, C and Ceresoli, D and Cococcioni, M and Colonna, N and Carnimeo, I and {Dal Corso}, A and {De Gironcoli}, S. and Delugas, P and Distasio, R. A. and Ferretti, A and Floris, A and Fratesi, G and Fugallo, G and Gebauer, R and Gerstmann, U and Giustino, F and Gorni, T and Jia, J and Kawamura, M and Ko, H. Y. and Kokalj, A and K{\"{u}}c{\"{u}}kbenli, E. and Lazzeri, M and Marsili, M and Marzari, N and Mauri, F and Nguyen, N. L. and Nguyen, H. V. and Otero-De-La-Roza, A. and Paulatto, L and Ponc{\'{e}}, S and Rocca, D and Sabatini, R and Santra, B and Schlipf, M and Seitsonen, A P and Smogunov, A and Timrov, I and Thonhauser, T and Umari, P and Vast, N and Wu, X and Baroni, S},
	doi = {10.1088/1361-648X/aa8f79},
	issn = {1361648X},
	journal = {J. Phys. Condens. Matter},
	month = {nov},
	number = {46},
	pages = {465901},
	title = {{Advanced} capabilities for materials modelling with {Quantum ESPRESSO}},
	url = {https://iopscience.iop.org/article/10.1088/1361-648X/aa8f79},
	volume = {29},
	year = {2017}
}

@article{Gladh2013,
title = {Electron- and phonon-coupling in femtosecond laser-induced desorption of {CO} from {Ru(0001)}},
journal = {Surface Science},
volume = {615},
pages = {65-71},
year = {2013},
issn = {0039-6028},
doi = {https://doi.org/10.1016/j.susc.2013.05.002},
url = {https://www.sciencedirect.com/science/article/pii/S0039602813001404},
author = {J. Gladh and T. Hansson and H. {\"{O}}ström},
}

@article{Gonzalez2025,
    author = {Gonzalez, Federico J. and Muzas, Alberto S. and Juaristi, J. Iñaki and Alducin, Maite and Busnengo, H. Fabio},
    title = {Femtosecond laser-induced diffusion and desorption of CO adsorbed on a weak electron–phonon coupling surface: Cu(110)},
    journal = {J. Chem. Phys.},
    volume = {162},
    number = {17},
    pages = {174701},
    year = {2025},
    month = {05},
    issn = {0021-9606},
    doi = {10.1063/5.0256832},
    url = {https://doi.org/10.1063/5.0256832}
}

@article{Guo2018,
author = {Guo, Chenxi and Wang, Ziyun and Wang, Dong and Wang, Hai-Feng and Hu, P.},
title = {First-Principles Determination of {CO} Adsorption and Desorption on {Pt(111)} in the Free Energy Landscape},
journal = {J. Phys. Chem. C},
volume = {122},
number = {37},
pages = {21478-21483},
year = {2018},
doi = {10.1021/acs.jpcc.8b06782},
url = {https://doi.org/10.1021/acs.jpcc.8b06782}
}

@article{Guo99,
title = {Theory of photoinduced surface reactions of admolecules},
journal = {Prog. Surf. Sci.},
volume = {62},
number = {7},
pages = {239-303},
year = {1999},
issn = {0079-6816},
doi = {https://doi.org/10.1016/S0079-6816(99)00013-1},
url = {https://www.sciencedirect.com/science/article/pii/S0079681699000131},
author = {Hua Guo and Peter Saalfrank and Tamar Seideman}
}

@article{Hayashi2013,
  title = {High-resolution angle-resolved photoemission study of electronic structure and electron self-energy in palladium},
  author = {Hayashi, Hirokazu and Shimada, Kenya and Jiang, Jian and Iwasawa, Hideaki and Aiura, Yoshihiro and Oguchi, Tamio and Namatame, Hirofumi and Taniguchi, Masaki},
  journal = {Phys. Rev. B},
  volume = {87},
  issue = {3},
  pages = {035140},
  numpages = {8},
  year = {2013},
  month = {Jan},
  publisher = {American Physical Society},
  doi = {10.1103/PhysRevB.87.035140},
  url = {https://link.aps.org/doi/10.1103/PhysRevB.87.035140}
}

@article{Hellsing1984,
doi = {10.1088/0031-8949/29/4/014},
url = {https://dx.doi.org/10.1088/0031-8949/29/4/014},
year = {1984},
month = {apr},
publisher = {},
volume = {29},
number = {4},
pages = {360},
author = {B Hellsing and  M Persson},
title = {Electronic Damping of Atomic and Molecular Vibrations at Metal Surfaces},
journal = {Phys. Scr.}
}

@Article{Hong2016,
  author  = {Hong, Sung-Young and Xu, Pan and Camillone, Nina R. and White, Michael G. and Camillone, Nicholas},
  journal = {J. Chem. Phys.},
  title   = {Adlayer Structure Dependent Ultrafast Desorption Dynamics in Carbon Monoxide Adsorbed on {Pd} (111)},
  year    = {2016},
  number  = {1},
  pages   = {014704},
  volume  = {145},
  doi     = {10.1063/1.4954408},
}

@article{juaristi08,
title = {Role of Electron-Hole Pair Excitations in the Dissociative Adsorption of Diatomic Molecules on Metal Surfaces},
author = {Juaristi, J. I. and Alducin, M.  and D\'{\i}ez Mui\~no, R. and Busnengo, H. F. and Salin, A. },
journal = {Phys. Rev. Lett.},
volume = {100},
number = {11},
pages = {116102},
numpages = {4},
year = {2008},
month = {Mar},
doi = {10.1103/PhysRevLett.100.116102},
url = {https://link.aps.org/doi/10.1103/PhysRevLett.100.116102},
publisher = {American Physical Society}
}

@Article{juaristiprb17,
  author    = {Juaristi, J. I{\~n}aki and Alducin, Maite and Saalfrank, Peter},
  title     = {Femtosecond Laser Induced Desorption of {${\mathrm{H}}_{2},{\mathrm{D}}_{2}$}, and {HD} from {Ru(0001)}: Dynamical Promotion and Suppression Studied with Ab Initio Molecular Dynamics with Electronic Friction},
  doi       = {10.1103/PhysRevB.95.125439},
  pages     = {125439},
  url       = {https://link.aps.org/doi/10.1103/PhysRevB.95.125439},
  volume    = {95},
  journal   = {Phys. Rev. B},
  month     = {Mar},
  publisher = {American Physical Society},
  year      = {2017},
}

@article{Kao93,
  title = {Femtosecond laser desorption of molecularly adsorbed oxygen from {Pt(111)}},
  author = {Kao, F.-J. and Busch, D. G. and Cohen, D. and Gomes da Costa, D. and Ho, W.},
  journal = {Phys. Rev. Lett.},
  volume = {71},
  issue = {13},
  pages = {2094--2097},
  numpages = {0},
  year = {1993},
  month = {Sep},
  publisher = {American Physical Society},
  doi = {10.1103/PhysRevLett.71.2094},
  url = {https://link.aps.org/doi/10.1103/PhysRevLett.71.2094}
}

@book{Kittel1986,
author  = {C. Kittel},
title   = {Introduction to solid state physics},
publisher  = {Wiley \& Sons, New York},
edition = {6th edition},
year    = 1986
}

@article{Kratzer2022,
  title = {Relaxation of photoexcited hot carriers beyond multitemperature models: General theory description verified by experiments on {Pb/Si(111)}},
  author = {Kratzer, Peter and Rettig, Laurenz and Sklyadneva, Irina Yu. and Chulkov, Evgueni V. and Bovensiepen, Uwe},
  journal = {Phys. Rev. Res.},
  volume = {4},
  issue = {3},
  pages = {033218},
  numpages = {12},
  year = {2022},
  month = {Sep},
  publisher = {American Physical Society},
  doi = {10.1103/PhysRevResearch.4.033218},
  url = {https://link.aps.org/doi/10.1103/PhysRevResearch.4.033218}
}

@article{Li2022,
title = {Ab initio calculation of electron temperature dependent electron heat capacity and electron-phonon coupling factor of noble metals},
journal = {Comp. Mater. Sci.},
volume = {202},
pages = {110959},
year = {2022},
issn = {0927-0256},
doi = {https://doi.org/10.1016/j.commatsci.2021.110959},
url = {https://www.sciencedirect.com/science/article/pii/S0927025621006558},
author = {Yongnan Li and Pengfei Ji}
}

@article{Lindner2023,
author = {Lindner, Steven and Lon\ifmmode \check{c}\else \v{c}\fi{}ari\ifmmode \acute{c}\else \'{c}\fi{}, Ivor and Vr\ifmmode \check{c}\else \v{c}\fi{}ek, Lovro and Alducin, Maite and Juaristi, J. I{\~n}aki and Saalfrank, Peter},
title = {Femtosecond Laser-Induced Desorption of Hydrogen Molecules from Ru(0001): A Systematic Study Based on Machine-Learned Potentials},
journal = {J. Phys. Chem. C},
volume = {127},
number = {30},
pages = {14756-14764},
year = {2023},
doi = {10.1021/acs.jpcc.3c02941},
url = {https://doi.org/10.1021/acs.jpcc.3c02941}
}

@Article{Lisowski2004,
author={Lisowski, M.
and Loukakos, P. A.
and Bovensiepen, U.
and St{\"a}hler, J.
and Gahl, C.
and Wolf, M.},
title={Ultra-fast dynamics of electron thermalization, cooling and transport effects in {Ru(001)}},
journal={Appl. Phys. A},
year={2004},
month={Jan},
day={01},
volume={78},
number={2},
pages={165-176},
issn={1432-0630},
doi={10.1007/s00339-003-2301-7},
url={https://doi.org/10.1007/s00339-003-2301-7}
}

@Article{loncaricprb16,
  author    = {Lon\ifmmode \check{c}\else \v{c}\fi{}ari\ifmmode \acute{c}\else \'{c}\fi{}, Ivor and Alducin, M. and Saalfrank, P. and Juaristi, J. I.},
  title     = {Femtosecond-Laser-Driven Molecular Dynamics on Surfaces: Photodesorption of Molecular {Oxygen} from {Ag(110)}},
  doi       = {10.1103/PhysRevB.93.014301},
  pages     = {014301},
  url       = {http://link.aps.org/doi/10.1103/PhysRevB.93.014301},
  volume    = {93},
  journal   = {Phys. Rev. B.},
  month     = {Jan},
  publisher = {American Physical Society},
  year      = {2016},
}

@Article{loncaricnimb16,
  author  = {Ivor Lon\v{c}ari\'{c} and Maite Alducin and Peter Saalfrank and J. Inaki Juaristi},
  title   = {Femtosecond Laser Pulse Induced Desorption: a Molecular Dynamics Simulation},
  doi     = {http://dx.doi.org/10.1016/j.nimb.2016.02.051},
  issn    = {0168-583X},
  pages   = {114 - 118},
  url     = {http://www.sciencedirect.com/science/article/pii/S0168583X16001816},
  volume  = {382},
  journal = {Nucl. Instrum. Methods B},
  year    = {2016},
}

@article{Luntz2006,
	author={Luntz, A. C. and Persson, M. and Wagner, S. and Frischkorn, C. and Wolf, M.},
    journal={J. Chem. Phys.},
    title={Femtosecond Laser Induced Associative Desorption of {H2} from {Ru(0001)}: Comparison of {\textquotedblleft}First Principles{\textquotedblright} Theory with Experiment},
    year={2006},
    volume={124},
    number = {24},
    pages={244702},
    month = {06},
    issn = {0021-9606},
    doi = {10.1063/1.2206588},
    url = {https://doi.org/10.1063/1.2206588}
}

@article{Miiller1971,
  author       = {A. P. Miiller and B. N. Brockhouse},
  journal      = {Can. J. Phys.},
  title        = {Crystal Dynamics and Electronic Specific Heats of Palladium and Copper},
  doi          = {10.1139/p71-087},
  number       = {6},
  pages        = {704--723},
  volume       = {49},
  year         = 1971
}

@article{Misewich94,
    author = {Misewich, J. A. and Kalamarides, A. and Heinz, T. F. and H{\"{o}}fer, U. and Loy, M. M. T.},
    title = {Vibrationally assisted electronic desorption: Femtosecond surface chemistry of {O2/Pd(111)}},
    journal = {J. Chem. Phys.},
    volume = {100},
    number = {1},
    pages = {736-739},
    year = {1994},
    month = {01},
    issn = {0021-9606},
    doi = {10.1063/1.466941},
    url = {https://doi.org/10.1063/1.466941},
}

@article{Mladineo2025,
    author = {Mladineo, Bruno and Juaristi, J. I{\~{n}}aki and Alducin, Maite and Saalfrank, Peter and Lon{\v{c}}ari{\'{c}}, Ivor},
    title = {Photoinduced dynamics of CO on Ru(0001): Understanding experiments by simulations with all degrees of freedom},
    journal = {J. Chem. Phys.},
    volume = {163},
    number = {4},
    pages = {044103},
    year = {2025},
    month = {07},
    issn = {0021-9606},
    doi = {10.1063/5.0278850},
    url = {https://doi.org/10.1063/5.0278850}
}

@article{Muzas2024,
author = {S. Muzas, Alberto and Serrano Jim{\'e}nez, Alfredo and Zhang, Yaolong and Jiang, Bin and Juaristi, J. I{\~n}aki and Alducin, Maite},
title = {Multicoverage Study of Femtosecond Laser-Induced Desorption of {CO} from {Pd(111)}},
journal = {J. Phys. Chem. Lett.},
volume = {15},
number = {9},
pages = {2587-2594},
year = {2024},
doi = {10.1021/acs.jpclett.4c00026},
url = {https://doi.org/10.1021/acs.jpclett.4c00026}
}

@article{NOFFSINGER20102140,
title = {{EPW}: A program for calculating the electron–phonon coupling using maximally localized  {Wannier} functions},
journal = {Comput. Phys. Commun.},
volume = {181},
number = {12},
pages = {2140-2148},
year = {2010},
issn = {0010-4655},
doi = {https://doi.org/10.1016/j.cpc.2010.08.027},
url = {https://www.sciencedirect.com/science/article/pii/S0010465510003218},
author = {Jesse Noffsinger and Feliciano Giustino and Brad D. Malone and Cheol-Hwan Park and Steven G. Louie and Marvin L. Cohen}
}

@article{Novko2016,
  title = {Effects of electronic relaxation processes on vibrational linewidths of adsorbates on surfaces: The case of {CO}/{Cu}(100)},
  author = {Novko, D. and Alducin, M. and Blanco-Rey, M. and Juaristi, J. I.},
  journal = {Phys. Rev. B},
  volume = {94},
  issue = {22},
  pages = {224306},
  numpages = {19},
  year = {2016},
  month = {Dec},
  publisher = {American Physical Society},
  doi = {10.1103/PhysRevB.94.224306},
  url = {https://link.aps.org/doi/10.1103/PhysRevB.94.224306}
}

@article{Novko2018,
  title = {Electron-Mediated Phonon-Phonon Coupling Drives the Vibrational Relaxation of {CO} on {Cu}(100)},
  author = {Novko, D. and Alducin, M. and Juaristi, J. I.},
  journal = {Phys. Rev. Lett.},
  volume = {120},
  issue = {15},
  pages = {156804},
  numpages = {6},
  year = {2018},
  month = {Apr},
  publisher = {American Physical Society},
  doi = {10.1103/PhysRevLett.120.156804},
  url = {https://link.aps.org/doi/10.1103/PhysRevLett.120.156804}
}

@article{Oberg2015,
    author = {{\"{O}}berg, H. and Gladh, J. and Marks, K. and Ogasawara, H. and Nilsson, A. and Pettersson, L. G. M. and {\"{O}}str{\"{o}}m, H.},
    title = {Indication of non-thermal contribution to visible femtosecond laser-induced {CO} oxidation on {Ru(0001})},
    journal = {J. Chem. Phys.},
    volume = {143},
    number = {7},
    pages = {074701},
    year = {2015},
    month = {08},
    issn = {0021-9606},
    doi = {10.1063/1.4928646},
    url = {https://doi.org/10.1063/1.4928646}
}

@article{PONCE2016116,
title = "{EPW}: Electron–phonon coupling, transport and superconducting properties using maximally localized  {Wannier} functions",
journal = "Comput. Phys. Commun.",
volume = "209",
pages = "116 - 133",
year = "2016",
issn = "0010-4655",
doi = "https://doi.org/10.1016/j.cpc.2016.07.028",
url = "http://www.sciencedirect.com/science/article/pii/S0010465516302260",
author = "S. Poncé and E.R. Margine and C. Verdi and F. Giustino"
}

@article{Prybyla1992,
  title = "{Femtosecond time-resolved surface reaction: Desorption of {CO} from Cu(111) in $<$325 fsec}",
  author = {Prybyla, J. A. and Tom, H. W. K. and Aumiller, G. D.},
  journal = {Phys. Rev. Lett.},
  volume = {68},
  issue = {4},
  pages = {503--506},
  numpages = {0},
  year = {1992},
  month = {Jan},
  publisher = {American Physical Society},
  doi = {10.1103/PhysRevLett.68.503},
  url = {https://link.aps.org/doi/10.1103/PhysRevLett.68.503}
}

@Article{saalfrank2006,
  author               = {Saalfrank, Peter},
  title                = {Quantum dynamical approach to ultrafast molecular desorption from surfaces},
  journal              = {Chem. Rev.},
  year                 = {2006},
  volume               = {106},
  number               = {10},
  month                = {OCT 11},
  pages                = {4116-4159},
  issn                 = {0009-2665},
  doi                  = {10.1021/cr0501691},
  eissn                = {1520-6890},
  orcid-numbers        = {Saalfrank, Peter/0000-0002-5988-5945},
  researcherid-numbers = {Saalfrank, Peter/AAT-6961-2020},
  unique-id            = {ISI:000241157200002}
}

@Article{Scholz2016,
  title = {Femtosecond-laser induced dynamics of {CO} on {Ru(0001)}: Deep insights from a hot-electron friction model including surface motion},
  author = {Scholz, Robert and Flo\ss{}, Gereon and Saalfrank, Peter and F\"uchsel, Gernot and Lon\ifmmode \check{c}\else \v{c}\fi{}ari\ifmmode \acute{c}\else \'{c}\fi{}, Ivor and Juaristi, J. I.},
  journal = {Phys. Rev. B},
  volume = {94},
  issue = {16},
  pages = {165447},
  numpages = {17},
  year = {2016},
  month = {Oct},
  publisher = {American Physical Society},
  doi = {10.1103/PhysRevB.94.165447},
  url = {https://link.aps.org/doi/10.1103/PhysRevB.94.165447}
}

@Article{Scholz2019,
  title = {Vibrational response and motion of carbon monoxide on {Cu(100)} driven by femtosecond laser pulses: Molecular dynamics with electronic friction},
  author = {Scholz, Robert and Lindner, Steven and Lon\ifmmode \check{c}\else \v{c}\fi{}ari\ifmmode \acute{c}\else \'{c}\fi{}, Ivor and Tremblay, Jean Christophe and Juaristi, J. I{\~n}aki and Alducin, M. and Saalfrank, Peter},
  journal = {Phys. Rev. B},
  volume = {100},
  issue = {24},
  pages = {245431},
  numpages = {20},
  year = {2019},
  month = {Dec},
  publisher = {American Physical Society},
  doi = {10.1103/PhysRevB.100.245431},
  url = {https://link.aps.org/doi/10.1103/PhysRevB.100.245431}
}

@Article{serrano2021,
  author    = {Alfredo Serrano-Jim{\'{e}}nez and Alberto P. S{\'{a}}nchez Muzas and Yaolong Zhang and Juraj Ov{\v{c}}ar and Bin Jiang and Ivor Lon{\v{c}}ari{\'{c}} and J. I{\~{n}}aki Juaristi and Maite Alducin},
  title     = {Photoinduced Desorption Dynamics of {CO} from {Pd(111)}: A Neural Network Approach},
  doi       = {10.1021/acs.jctc.1c00347},
  number    = {8},
  pages     = {4648--4659},
  url       = {https://doi.org/10.1021%2Facs.jctc.1c00347},
  volume    = {17},
  journal   = {J. Chem. Theory Comput.},
  month     = {jul},
  publisher = {American Chemical Society ({ACS})},
  year      = {2021},
}

@Article{Spiering2018,
  author  = {Spiering, Paul and Meyer, J{\"o}rg},
  journal = {J. Phys. Chem. Lett.},
  title   = {Testing Electronic Friction Models: Vibrational De-excitation in Scattering of H2 and D2 from Cu(111)},
  year    = {2018},
  number  = {7},
  pages   = {1803-1808},
  volume  = {9},
  doi     = {10.1021/acs.jpclett.7b03182},
  url     = {https://doi.org/10.1021/acs.jpclett.7b03182},
}

@article{Springer1994, 
    author={Springer, C. and Head-Gordon, M. and Tully, J. C.},
    journal={Surf. Sci.},
    title = {Simulations of Femtosecond Laser-Induced Desorption of {CO} from {Cu}(100)},
    year={1994},
    volume={320},
    pages={L57--L62},
}

@article{Springer1996,
	 author={Springer, C. and Head-Gordon, M.},
    journal={Chem. Phys.},
    title={Simulations of the Femtosecond Laser-Induced Desorption of {CO} from {Cu}(100) at 0.5 {ML} Coverage},
    year={1996},
    volume={205},
    pages={73--89},
}

@article{Schendel2017,
  title = {Strong paramagnon scattering in single atom {Pd} contacts},
  author = {Schendel, V. and Barreteau, C. and Brandbyge, M. and Borca, B. and Pentegov, I. and Schlickum, U. and Ternes, M. and Wahl, P. and Kern, K.},
  journal = {Phys. Rev. B},
  volume = {96},
  issue = {3},
  pages = {035155},
  numpages = {5},
  year = {2017},
  month = {Jul},
  publisher = {American Physical Society},
  doi = {10.1103/PhysRevB.96.035155},
  url = {https://link.aps.org/doi/10.1103/PhysRevB.96.035155}
}

@article{Struck96,
  title = {Femtosecond Laser-Induced Desorption of {CO} from {Cu(100)}: Comparison of Theory and Experiment},
  author = {Struck, Lisa M. and Richter, Lee J. and Buntin, Steven A. and Cavanagh, Richard R. and Stephenson, John C.},
  journal = {Phys. Rev. Lett.},
  volume = {77},
  issue = {22},
  pages = {4576--4579},
  numpages = {0},
  year = {1996},
  month = {Nov},
  publisher = {American Physical Society},
  doi = {10.1103/PhysRevLett.77.4576},
  url = {https://link.aps.org/doi/10.1103/PhysRevLett.77.4576}
}

@article{Szymanski2007,
title = {Temperature-dependent electron-mediated coupling in subpicosecond photoinduced desorption},
journal = {Surf. Sci.},
volume = {601},
number = {16},
pages = {3335-3349},
year = {2007},
issn = {0039-6028},
doi = {https://doi.org/10.1016/j.susc.2007.06.004},
url = {https://www.sciencedirect.com/science/article/pii/S0039602807006395},
author = {P. Szymanski and A.L. Harris and N. Camillone},
}

@article{tetenoire2022,
author = {Tetenoire, Auguste and Ehlert, Christopher and Juaristi, J. I{\~n}aki and Saalfrank, Peter and Alducin, M.},
title = {Why Ultrafast Photoinduced {CO} Desorption Dominates over Oxidation on {Ru(0001)}},
journal = {J. Phys. Chem. Lett.},
volume = {13},
number = {36},
pages = {8516-8521},
year = {2022},
doi = {10.1021/acs.jpclett.2c02327},
url = {https://doi.org/10.1021/acs.jpclett.2c02327}
}

@article{tetenoire2023,
author = {Tetenoire, Auguste and Juaristi, J. I{\~n}aki and Alducin, Maite},
title = {{Photoinduced {CO} Desorption Dominates over Oxidation on Different {O +CO} Covered{ Ru(0001) }Surfaces}},
journal = {J. Phys. Chem. C},
year    = {2023},
number  = {21},
pages   = {10087-10096},
volume  = {127},
doi = {10.1021/acs.jpcc.3c01192},
url = {https://doi.org/10.1021/acs.jpcc.3c01192}
}

@article{Vazhappilly2009,
  author               = {Vazhappilly, Tijo and Klamroth, Tillmann and Saalfrank, Peter and Hernandez, Rigoberto},
  date                 = {2009-05},
  year                 = {2009},
  journal              = {J. Phys. Chem. C},
  title                = {{Femtosecond-Laser Desorption of {H-2 (D-2)} from {Ru(0001)}: Quantum and Classical Approaches}},
  doi                  = {10.1021/jp810709k},
  issn                 = {1932-7447},
  number               = {18},
  pages                = {7790-7801},
  volume               = {113},
  eissn                = {1932-7455},
  publisher            = {American Chemical Society ({ACS})},
url = {https://doi.org/10.1021/jp810709k}
}

@article{Wagner2008,
doi = {10.1088/1367-2630/10/12/125031},
url = {https://dx.doi.org/10.1088/1367-2630/10/12/125031},
year = {2008},
month = {dec},
publisher = {},
volume = {10},
number = {12},
pages = {125031},
author = {Wagner, S and {\"{O}}ström, H and Kaebe, A and Krenz, M and Wolf, M and Luntz, A C and Frischkorn, C},
title = {Activated associative desorption of {C + O →CO} from {Ru(001)} induced by femtosecond laser pulses},
journal = {New J. Phys.}
}

@article{Wang2025,
    author = {Wang, Xiangrui and Wang, Jiamin and Spiering, Paul and Liu, Liping and Meyer, Jörg and LaRue, Jerry L. and Xin, Hongliang},
    title = {Unveiling the interplay of electronic and phononic excitations in laser-induced oxygen activation on Ru(0001)},
    journal = {J. Chem. Phys.},
    volume = {163},
    number = {11},
    pages = {114109},
    year = {2025},
    month = {09},
    issn = {0021-9606},
    doi = {10.1063/5.0278197},
    url = {https://doi.org/10.1063/5.0278197}
}

@article{winter2003,
  title = {Energy loss of slow ions in a nonuniform electron gas},
  author = {Winter, H. and Juaristi, J. I. and Nagy, I. and Arnau, A. and Echenique, P. M.},
  journal = {Phys. Rev. B},
  volume = {67},
  issue = {24},
  pages = {245401},
  numpages = {6},
  year = {2003},
  month = {Jun},
  publisher = {American Physical Society},
  doi = {10.1103/PhysRevB.67.245401},
  url = {https://link.aps.org/doi/10.1103/PhysRevB.67.245401}
}

@article{yamanaka2002,
title = {State-resolved femtosecond two-pulse correlation measurements of {NO} photodesorption from {Pt(111)}},
journal = {Surface Science},
volume = {514},
number = {1},
pages = {404-408},
year = {2002},
issn = {0039-6028},
doi = {https://doi.org/10.1016/S0039-6028(02)01659-X},
url = {https://www.sciencedirect.com/science/article/pii/S003960280201659X},
author = {T. Yamanaka and A. Hellman and Shiwu Gao and W. Ho}
}

@article{Zhang2021,
  title = {Finite-temperature density-functional-theory investigation on the nonequilibrium transient warm-dense-matter state created by laser excitation},
  author = {Zhang, Hengyu and Zhang, Shen and Kang, Dongdong and Dai, Jiayu and Bonitz, M.},
  journal = {Phys. Rev. E},
  volume = {103},
  issue = {1},
  pages = {013210},
  numpages = {11},
  year = {2021},
  month = {Jan},
  publisher = {American Physical Society},
  doi = {10.1103/PhysRevE.103.013210},
  url = {https://link.aps.org/doi/10.1103/PhysRevE.103.013210}
}

@article{Zugec2024,
author = {Žugec, Ivan and Tetenoire, Auguste and Muzas, Alberto S. and Zhang, Yaolong and Jiang, Bin and Alducin, Maite and Juaristi, J. I{\~n}aki},
title = {Understanding the Photoinduced Desorption and Oxidation of {CO} on {Ru(0001)} Using a Neural Network Potential Energy Surface},
journal = {JACS Au},
volume = {4},
number = {5},
pages = {1997-2004},
year = {2024},
doi = {10.1021/jacsau.4c00197},
url = {https://doi.org/10.1021/jacsau.4c00197}
}

@book{Lide2000,
editor={Lide, DR},
author = {Lide, DR},
title={CRC Handbook of Chemistry and Physics},
year ={2000},
pages = {4-77},
publisher = {CRC Press LLC},
edition = {81},
address = {Boca Ranton: FL}
}

@article{Smirnov2025,
  title = {Electronic structure, lattice dynamics, and electron-phonon coupling factor of metals under nonequilibrium heating},
  author = {Smirnov, N. A.},
  journal = {Phys. Rev. B},
  volume = {111},
  issue = {1},
  pages = {014107},
  numpages = {14},
  year = {2025},
  month = {Jan},
  publisher = {American Physical Society},
  doi = {10.1103/PhysRevB.111.014107},
  url = {https://link.aps.org/doi/10.1103/PhysRevB.111.014107}
}

@article{Akhmetov2023,
  title = {Electron-phonon coupling in transition metals beyond Wang's approximation},
  author = {Akhmetov, Fedor and Milov, Igor and Makhotkin, Igor A. and Ackermann, Marcelo and Vorberger, Jan},
  journal = {Phys. Rev. B},
  volume = {108},
  issue = {21},
  pages = {214301},
  numpages = {11},
  year = {2023},
  month = {Dec},
  publisher = {American Physical Society},
  doi = {10.1103/PhysRevB.108.214301},
  url = {https://link.aps.org/doi/10.1103/PhysRevB.108.214301}
}

@article{Medvedev2024,
title = {Electronic heat conductivity in a two-temperature state},
journal = {International Journal of Heat and Mass Transfer},
volume = {228},
pages = {125674},
year = {2024},
issn = {0017-9310},
doi = {https://doi.org/10.1016/j.ijheatmasstransfer.2024.125674},
url = {https://www.sciencedirect.com/science/article/pii/S0017931024005052},
author = {Nikita Medvedev and Fedor Akhmetov and Igor Milov}
}

\end{document}